\documentclass[aps,prd,amsmath,superscriptaddress,%
               nobibnotes,nofootinbib,preprintnumbers,
               onecolumn,12pt,tightenlines]{revtex4}

\usepackage{graphicx}

\usepackage{ifthen}

\usepackage{url}

\newcommand{\coloronlinestatement}{}

\newcommand{\eg}{\textit{e.g.}}
\newcommand{\etal}{\textit{et~al.}}

\newcommand{\reffig}[2][fig:]{Figure~\ref{#1#2}}
\newcommand{\Reffig}[2][fig:]{Figure~\ref{#1#2}}
\newcommand{\refsec}[2][sec:]{Section~\ref{#1#2}} 
\newcommand{\Refsec}[2][sec:]{Section~\ref{#1#2}} 

\newcommand{\ifmulticol}[2]{%
  \ifthenelse{\lengthtest{1.9\columnwidth<\textwidth}}{#1}{#2}%
}

\newcommand{\insertdoublefig}[3][0.45]{%
    \hspace*{\stretch{1}}
    \includegraphics[keepaspectratio,width=#1\textwidth]{#2}
    \includegraphics[keepaspectratio,width=#1\textwidth]{#3}
    \hspace*{\stretch{1}}
}

\newcommand{\gae}{%
  \ensuremath{\lower 2pt \hbox{%
    $\, \buildrel {\scriptstyle >}\over {\scriptstyle \sim}\,$}%
    }%
  }
\newcommand{\lae}{%
  \ensuremath{\lower 2pt \hbox{%
    $\, \buildrel {\scriptstyle <}\over {\scriptstyle \sim}\,$}%
    }%
  }

\newcommand{\tanb}{\ensuremath{\tan{\beta}}}

\newcommand{\rms}{\ensuremath{\mathrm{s}}}

\newcommand{\SigmapiN}{\ensuremath{\Sigma_{\pi\!{\scriptscriptstyle N}}}}

\newcommand{\Deltas}{\ensuremath{\Delta_{\rms}}}

\providecommand{\texorpdfstring}[2]{#1}

\begin{document}



\vspace*{-1.75cm}
\begin{minipage}[c]{0.2\textwidth}
  \begin{flushleft}
    \vspace{\baselineskip}
    \makebox[0pt][l]{
      \hspace*{-1.0cm}
      \includegraphics[keepaspectratio,width=2cm]%
                      {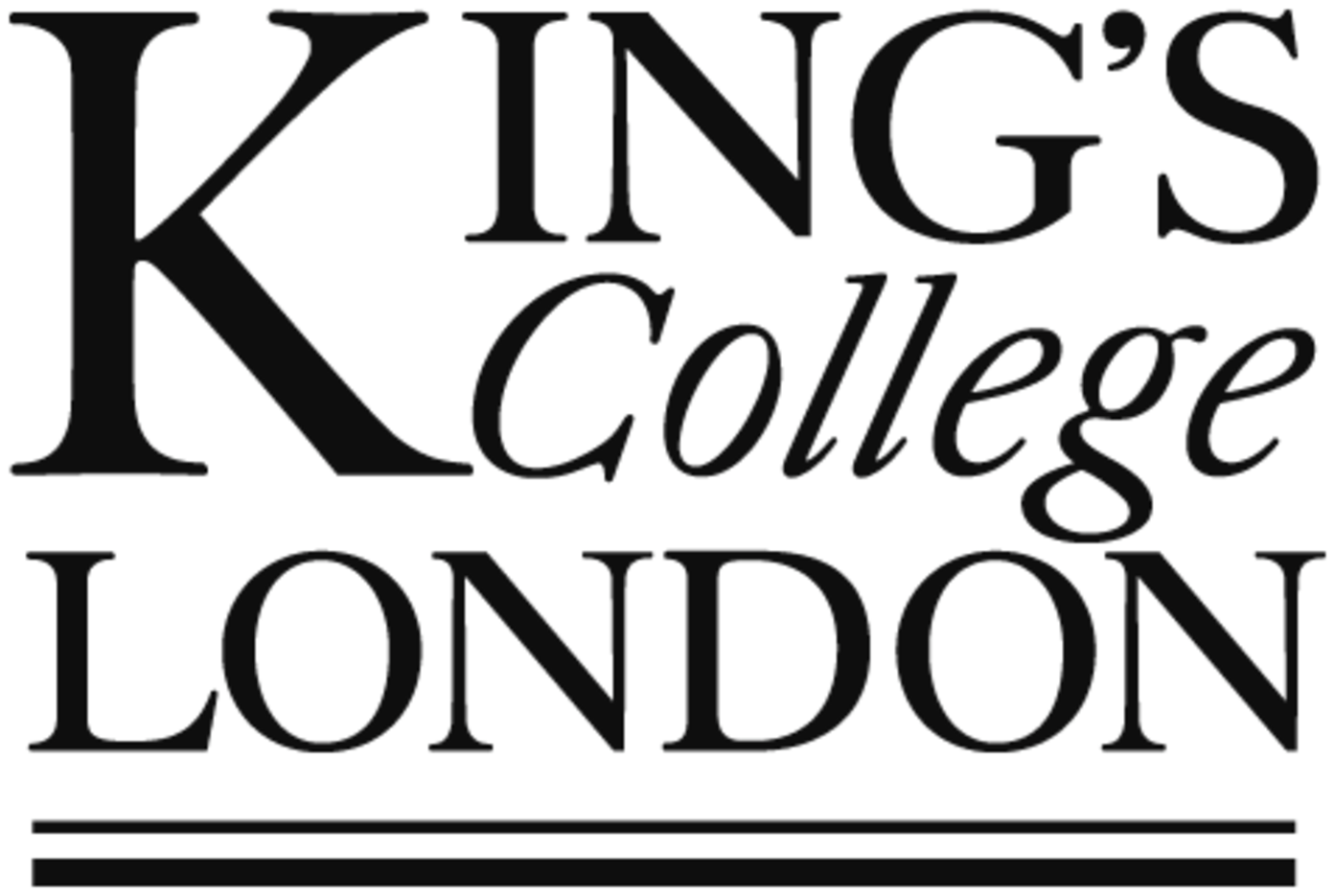}
    }
  \end{flushleft}
\end{minipage}
\hfill
\begin{minipage}[c]{0.79\textwidth}
  \begin{flushright}
    CERN-PH-TH/2011-025, KCL-PH-TH/2011-03, SU--ITP--11/04,\\
    UMN--TH--2936/11, FTPI--MINN--11/03, SLAC-PUB-14372
  \end{flushright}
\end{minipage}


\title{Neutrino Fluxes from NUHM LSP Annihilations in the Sun}

\author{\bf John Ellis}
\email{John.Ellis@cern.ch}
\affiliation{
 TH Division,
 Physics Department,
 CERN,
 1211 Geneva 23, Switzerland;\\
 Theoretical Physics and Cosmology Group,
 Department of Physics,
 King's College London,
 London WC2R 2LS, UK}

\author{\bf Keith A.\ Olive}
\email{olive@physics.umn.edu}
\affiliation{
 William I. Fine Theoretical Physics Institute,
 School of Physics and Astronomy,
 University of Minnesota,
 Minneapolis, MN 55455, USA;\\
 Department of Physics and SLAC,
 Stanford University,
 Palo Alto, CA 94305, USA}

\author{\bf Christopher Savage}
\email{savage@fysik.su.se}
\affiliation{
 The Oskar Klein Centre for Cosmoparticle Physics,
 Department of Physics,
 Stockholm University,
 AlbaNova,
 SE-10691 Stockholm, Sweden}

\author{\bf Vassilis C.\ Spanos}
\email{spanos@inp.demokritos.gr}
\affiliation{
 Institute of Nuclear Physics,
 NCSR ``Demokritos'',
 GR-15310 Athens, Greece}

\date{April 27, 2011}


\begin{abstract} 
%
%

{\small
We extend our previous studies of the neutrino fluxes expected from
neutralino LSP annihilations inside the Sun to include variants of the
minimal supersymmetric extension of the Standard Model (MSSM) with
squark, slepton and gaugino masses constrained to be universal at the
GUT scale, but allowing one or two non-universal supersymmetry-breaking
parameters contributing to the Higgs masses (NUHM1,2). As in the
constrained MSSM (CMSSM) with universal Higgs masses,
there are large regions of the NUHM parameter space where the LSP
density inside the Sun is not in equilibrium, 
so that the annihilation rate may be far below the capture rate,
and there are also large regions where the capture rate is
not dominated by spin-dependent LSP-proton scattering. 
The spectra possible in the NUHM are qualitatively similar to
those in the CMSSM. We calculate neutrino-induced muon
fluxes above a threshold energy of 10~GeV, appropriate for the 
IceCube/DeepCore detector, for points  
where the NUHM yields the correct cosmological relic density for
representative choices of the NUHM parameters.
We find that the IceCube/DeepCore detector can probe regions of the
NUHM parameter space in addition to analogues of the focus-point strip
and the tip of the coannihilation strip familiar from the CMSSM.
These include regions with enhanced Higgsino-gaugino mixing in the LSP
composition, that occurs where neutralino mass eigenstates cross over.
On the other hand, rapid-annihilation funnel regions in general yield
neutrino fluxes that are unobservably small.
}

\end{abstract} 

\maketitle


\section{\label{sec:intro} Introduction}

One of the most actively pursued strategies for detecting supersymmetric
dark matter particles (LSPs) is the search for signatures of the
annihilations of LSPs inside the Sun or Earth\cite{indirectdet:solar,
indirectdet:earth}. The principle for detection is to search for the
passage through a large detector in ice or water of muons produced by
the interactions of energetic neutrinos released in the LSP dark matter
annihilation process. There have been extensive studies of this potential
experimental signature in many variants of the MSSM, and experiments
such as IceCube/DeepCore~\cite{Ahrens:2003ix,Achterberg:2006md,
Resconi:2008fe,Abbasi:2009uz} are starting to chip away at the MSSM
parameter space.

We recently re-analyzed ~\cite{EOSSnu} this potential signature in the
framework of the MSSM with all the supersymmetry-breaking spin-1/2 and
-0 mass parameters $(m_{1/2}, m_0)$ constrained to be universal at the GUT
scale (the CMSSM) \cite{cmssm}, imposing the requirement that the LSP
should provide the density of dark matter inferred from WMAP
\cite{Komatsu:2010fb} and other experiments. In the CMSSM, where the
supersymmetry breaking trilinear mass parameters, $A_0$ are also taken
to be universal at the GUT scale, the resulting relic density is found
to lie in the WMAP range only along relatively narrow strips in the
$(m_{1/2}, m_0)$ plane for fixed $\tan \beta$ and $A_0$
\cite{eoss,cmssmwmap}.
These correspond to the coannihilation strip, where the mass of the
lightest neutralino is close to the mass of the lightest charged slepton
(usually the mostly right-handed stau); the heavy Higgs funnel, found at
large $\tan \beta$ and large $(m_{1/2}, m_0)$, where the neutralino mass
is close to half the heavy Higgs mass and rapid annihilations of 
neutralinos are mediated by the s-channel exchange of heavy Higgs
scalars and pseudoscalars; and the focus-point region which is typically
found at very large values of $m_0$ when the $\mu$ parameter (an output
of the minimization of the Higgs potential in the CMSSM) is driven to
small values and the neutralino picks up a more significant Higgsino
component. 

In our previous work \cite{EOSSnu}, we found that the LSP capture rate
was not in general dominated by scattering on protons inside the Sun via
spin-dependent couplings, but that an important role was
often played by spin-independent scattering on heavier nuclides.
We also found that, in many regions of the CMSSM parameter space, 
LSP capture and annihilation would not be in equilibrium, and that
the annihilation rate would be correspondingly reduced. We also
analyzed the uncertainties in the magnitude of the potential 
muon-neutrino signal due to uncertainties in the composition of the Sun
and in the scattering matrix elements. We found that the 
CMSSM might be detectable in IceCube/DeepCore along (some part of) the
WMAP strip in the focus-point region of the $(m_{1/2}, m_0)$ plane,
and near the low-$m_{1/2}$ tip of the WMAP strip in the 
coannihilation region \cite{cmssmnu,EOSSnu}.

In this paper we extend these previous studies to models with one or
two degrees of non-universality in the soft supersymmetry-breaking
contributions to the Higgs doublets, the NUHM1~\cite{nuhm1,nuhm12} and
NUHM2~\cite{nonu,nuhm2,nuhm12}. One of our
primary objectives is to understand the circumstances under which
such relatively high neutrino fluxes may be attained in these models
and, conversely, whether the relatively low rates usually
found in the CMSSM  are specific to that model. More generally,
we seek to lay a basis for systematic comparisons of the physics
capabilities of different detection strategies in (relatively) simple
variants of the MSSM.

Supersymmetric dark matter searches~\cite{Ahmed:2009zw,Aprile:2010um,
Abbasi:2009uz} are complementary to searches for
supersymmetry at accelerators~\cite{CMSsusy,ATLASsusy}. 
At the moment, the latter are sensitive
primarily to the spin-1/2 and -0 mass parameters $m_{1/2}$ and $m_0$,
and are less sensitive to the non-universality parameters 
that appear in the NUHM1,2.
For example, global likelihood fits \cite{mc3} currently yield similar
68 and 95\% confidence-level preferred regions in
the $(m_{1/2}, m_0)$ planes of the CMSSM and NUHM1. It is therefore
particularly interesting to know whether direct and indirect dark
matter searches offer ways to differentiate between these models.

The NUHM1,2 offer additional mechanisms to bring the relic LSP
density into the WMAP range, in addition to the coannihilation,
rapid-annihilation and
focus-point possibilities mentioned above in the context of the CMSSM.
For example, there are distinctive regions of NUHM parameter space where 
Higgsino-gaugino mixing in the LSP is enhanced by level-crossing
in the neutralino mass matrix, bringing the relic density into the WMAP
range. Alternatively, the LSPs may annihilate rapidly through
direct-channel heavy Higgs $H, A$ poles even if $m_{1/2}, m_0$ and
$\tanb$ are relatively small. It was shown in~\cite{mc3} that the
collider prospects for sparticle detection are rather different in the
NUHM1 low-mass rapid-annihilation region than they are in the CMSSM,
whereas the favoured rates for direct LSP detection via scattering on
nuclei were broadly similar in the CMSSM and the NUHM1 (though the
uncertainties were greater in the latter case). Therefore, it is
interesting to study the prospects in this region for LSP detection via
the energetic neutrinos produced by annihilation inside the Sun.

The layout of this paper is as follows. In \refsec{preamble} we recall
briefly some  general features of the NUHM1,2 and discuss other inputs
into the rate  calculations. Then, in \refsec{NUHM1} we explore the solar
annihilation rates in some generic slices through the NUHM1 parameter
space, finding that they are enhanced in regions with relatively large
Higgsino components in the LSP as may occur for specific relations between 
$\mu$ and $m_{1/2}$ where there is level crossing.
In \refsec{NUHM2} we extend our analysis to slices through the
NUHM2 parameter space. \Refsec{summary} summarizes our conclusions.

We find that whereas the neutrino flux may be observable in regions
with enhanced Higgsino-gaugino mixing, analogously to the enhancement
along the focus-point WMAP strip in the CMSSM, the flux is generically
unobservably low in the rapid-annihilation funnels. This suggests that
the observation of a high-energy solar neutrino flux in the
IceCube/DeepCore experiment is a potential diagnostic for large mixing
and level crossing in the neutralino mass matrix, and specifically of
the relation between $\mu$ and $m_{1/2}$, which is a potential tool for
identifying non-universal Higgs mass parameters.

\section{\label{sec:preamble} Preamble}

\subsection{The NUHM1 and NUHM2 Parameter Spaces}

In the CMSSM, the free parameters are the supposedly universal
supersymmetry-breaking parameters $m_{1/2}, m_0$ and $A_0$,
as well as $\tanb$. The Higgs mixing superpotential parameter $\mu$
and the bilinear supersymmetry-breaking parameter $B_0$, and
hence the pseudoscalar Higgs mass $m_A$, are then
determined using the electroweak vacuum conditions, with a sign
ambiguity in $\mu$. The sign of the discrepancy between the experimental
value of $g_\mu - 2$ and the value calculated within the Standard Model
suggests that $\mu > 0$, and CMSSM analyses are often presented in
$(m_{1/2}, m_0)$ planes for $\mu > 0$ and fixed values of $\tanb$
and $A_0$. Although much of our NUHM analysis is for $\mu > 0$, 
we also consider the possibility of a negative sign.
The value of $A_0$ is notoriously unconstrained,
see, \eg,~\cite{mcmSUGRA}, and for definiteness we set it to zero in
what follows.

In the NUHM1, the soft supersymmetry-breaking contributions to the
masses of the two MSSM Higgs doublets, $m_1$ and $m_2$, 
are assumed to be equal,
but are allowed to differ from $m_0$. The extra degree of freedom
may be used to treat either $m_A$ or $|\mu|$ as a free parameter. On the 
other hand, in the NUHM2, the two soft supersymmetry-breaking
contributions to the masses of the MSSM Higgs doublets are allowed to
vary independently, and the two extra degrees of freedom may be used to
treat both $m_A$ and $|\mu|$ as free parameters.
Analyses of the NUHM1 and NUHM2 parameter spaces are often presented in
planes spanned by pairs of the quantities $m_{1/2}, m_0, m_A$ and $\mu$
for some fixed values of $\tanb$ and the other parameters, 
and we present some examples below which are selected from the analysis
in \cite{nuhm12}. 
Since a frequentist likelihood analysis favours relatively small values
of $m_{1/2}$ and $m_0$ in both the CMSSM and NUHM1 \cite{mc3}, we
concentrate here on NUHM1 and NUHM2 planes with relatively low values of
either $m_{1/2}$ and $m_0$. We note, however, that a complete likelihood
analysis of the NUHM2 is yet to be performed.

\subsection{Spin-Dependent and -Independent Scattering Rates}

\begin{figure*}
  \insertdoublefig{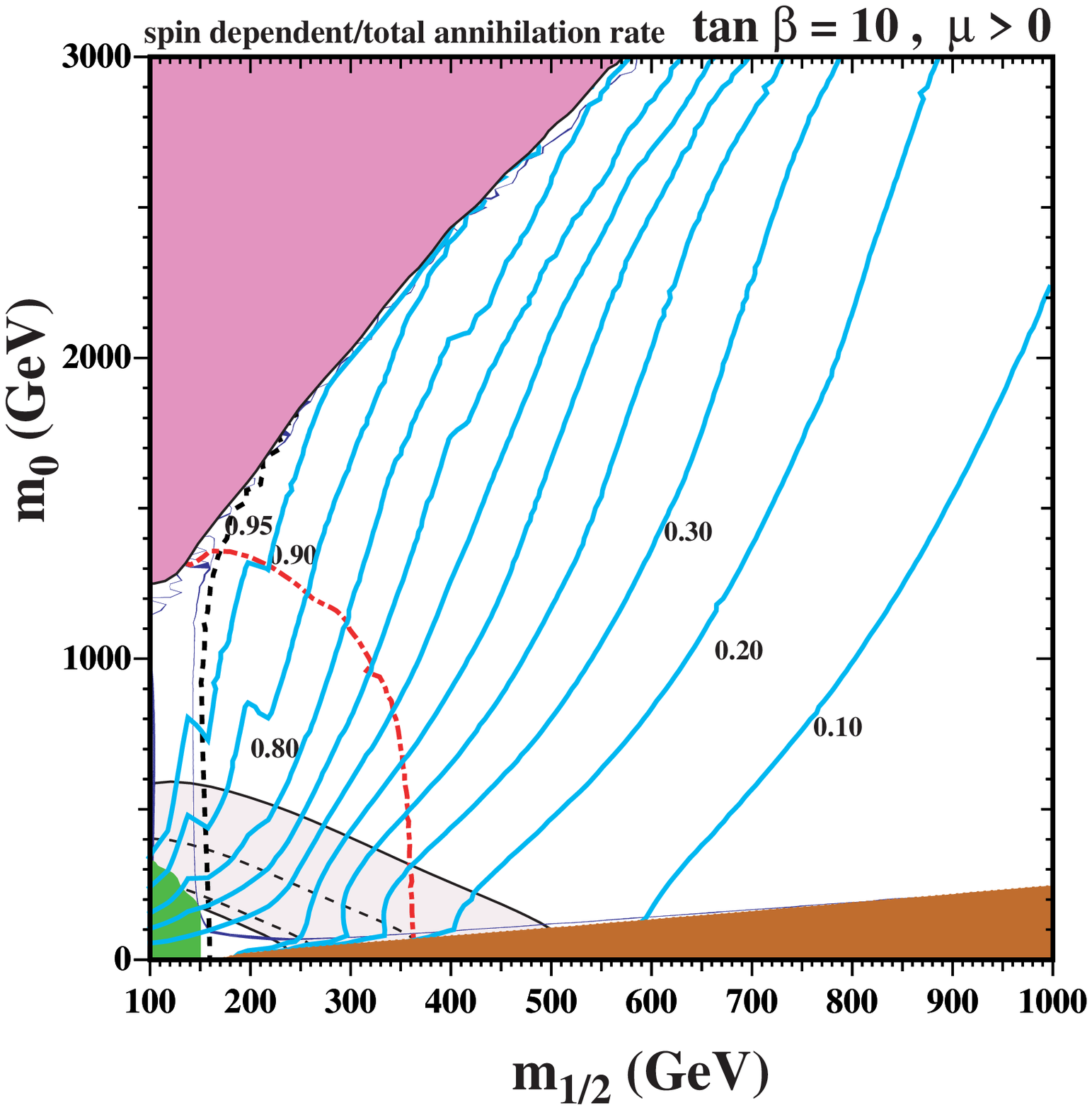}{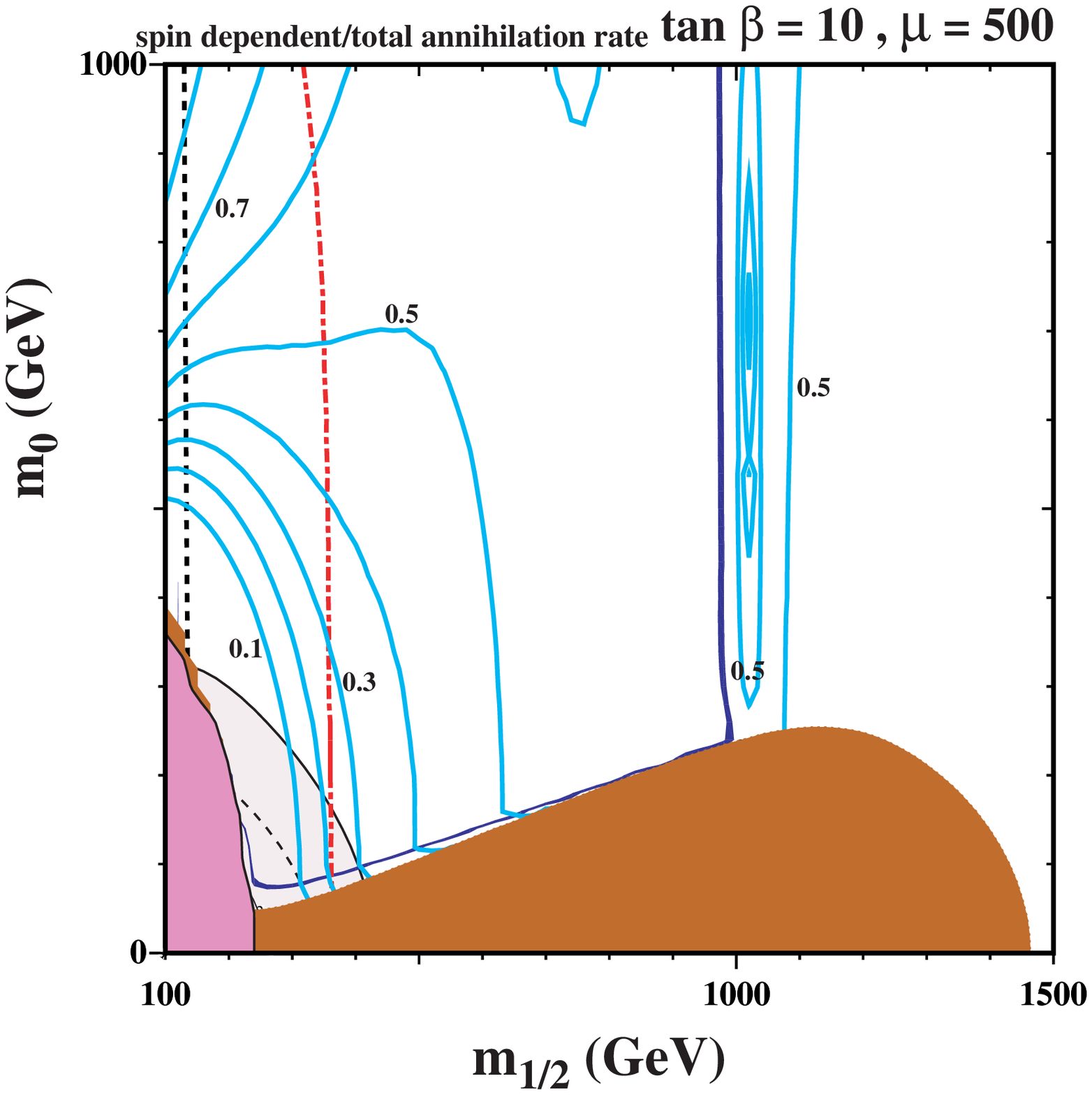}
  \caption{\it
    \coloronlinestatement
    Sample $(m_{1/2}, m_0)$ planes for $\tanb = 10$ and $A_0 = 0$ in
    (left) the CMSSM for $\mu > 0$ and
    (right) the NUHM1 for $\mu = 500$~GeV,
    showing regions excluded because there is no consistent electroweak
    vacuum (dark pink shading), 
    or because there is charged dark matter (brown shading),
    or because of a conflict with $b \to s \gamma$ measurements (green
    shading).  Only regions to the right of the black dashed (red
    dash-dotted) line are consistent with the absence at LEP of
    charginos (a Higgs boson).  The turquoise strips are favoured by the
    determination of the cold dark matter density by WMAP and other
    experiments \cite{Komatsu:2010fb}, and the light pink band is favoured
    by the BNL measurement of $g_\mu - 2$. We also show contours of the
    ratio of the solar dark matter annihilation rate calculated
    using only spin-dependent scattering to the total annihilation
    rate including also spin-independent scattering.
    }
  \label{fig:sds}
\end{figure*}

When calculating the LSP annihilation rates inside the Sun,
the key particle physics inputs---apart from the choice of
supersymmetric model---are the matrix elements for dark
matter scattering on the nuclides inside the Sun. It is often assumed
that LSP capture in the Sun is dominated by
spin-dependent scattering on hydrogen but, as discussed
in~\cite{EOSSnu}, spin-independent scattering on heavier
nuclei actually dominates in generic regions of the CMSSM
parameter space. \Reffig{sds} displays contours of the
ratio of the solar dark matter annihilation rate calculated 
using only spin-dependent scattering to the total annihilation
rate including also spin-independent scattering in (left) the
CMSSM $(m_{1/2}, m_0)$ plane for $\tan \beta = 10$, $A_0 = 0$
and $\mu > 0$, and (right) the NUHM1 $(m_{1/2}, m_0)$ plane for 
$\tan \beta = 10$, $A_0 = 0$ and $\mu = 500$~GeV.
Regions excluded because there is no consistent 
electroweak symmetry breaking (EWSB) have dark pink shading,
those with a charged LSP have brown shading,
and those in conflict with $b \to s \gamma$~\cite{bsgex} measurements
have green shading. Regions to the left of the black dashed (red
dash-dotted) line are inconsistent with the absence at LEP of
charginos (a Higgs boson) \cite{LEPsusy,LEPHiggs,FeynHiggs}.
We recall that there is a theoretical uncertainty $\sim 1.5$~GeV in the
calculation of $m_h$ in the CMSSM, which induces an uncertainty
$\sim 50$~GeV in values of $m_{1/2}$ along the red dash-dotted line.
The recent LHC direct exclusion regions in the CMSSM $(m_{1/2}, m_0)$
plane~\cite{CMSsusy,ATLASsusy} are somewhat weaker than the indirect LEP
Higgs constraint shown here. The pale pink band is favoured
by the BNL measurement of $g_\mu - 2$~\cite{Bennett:2006fi,g-2babar,newg-2},
at the $\pm 1 (2)-\sigma$ level along the dashed (solid) lines,
but we do not impose this as a constraint on our analysis. 
In the CMSSM case (left panel), we see that spin-dependent scattering is
dominant only at small $m_{1/2}$ and large $m_0$.
In the NUHM1 case (right panel), we see that spin-dependent scattering
is subdominant at small $m_{1/2}$, becoming more important along the
WMAP-compatible strip with $m_{1/2} \sim 1000$~GeV.
However, spin-independent scattering is important even here, and in the
following analysis all calculations of the annihilation rates include
the contributions from both spin-dependent and -independent scattering. 

The uncertainties in the spin-independent scattering matrix element
include those in the ratios of the light quark masses, the octet
contribution, $\sigma_0$, to the pion-nucleon $\sigma$ term, $\SigmapiN$,
and the value of $\SigmapiN$ itself \cite{Bottino:1999ei,Accomando:1999eg,
Ellis:2005mb,Ellis:2008hf,Niro:2009mw}. The largest uncertainty is due
to $\SigmapiN$, for which we used the value 64~MeV as our
default in~\cite{EOSSnu}, whilst also exploring the implications of
other values. The second-largest uncertainty is due to that in $\sigma_0$,
for which we used the value 36~MeV as our default. We assume the
same default values in this analysis.
As discussed in more detail in Refs.~\cite{Ellis:2008hf,EOSSnu},
different measurements for $\SigmapiN$ and $\sigma_0$ lead to
variations in the spin-independent scattering cross-section by a
factor of $\sim$2--3 and the choice of experimental values for these
two parameters significantly impacts the spin-independent scattering
contribution to capture in the Sun. If $\SigmapiN$ were smaller, the
total annihilation rate would decrease, and the ratio of the solar dark
matter annihilation rate calculated using only spin-dependent
scattering to the total annihilation rate including also
spin-independent scattering shown in \reffig{sds} would increase.
By comparison, the uncertainties
in the light quark mass ratios are much less significant.

The principal uncertainty in the spin-dependent matrix element is the
contribution of strange quarks to the nucleon spin, $\Deltas$, with the
uncertainties due to $g_A$ and the SU(3)-octet nucleon matrix elements
being significantly smaller. The range $-0.06 \ge \Deltas \ge -0.12$
was considered in~\cite{EOSSnu}, and was found to induce an uncertainty
in the annihilation rate that was $\sim 10$\%, considerably smaller than
the others considered.  In the following we take the central value
$\Deltas = - 0.09$~\cite{Alekseev:2007vi}, which was adopted
in~\cite{EOSSnu} as the default value.

\subsection{Capture/Annihilation Rates and Neutrino/Muon Fluxes}

We compared in~\cite{EOSSnu} the rates estimated in various alternative
solar models.
We found that there was at most a 4\% difference in the annihilation
rates between the two models of Serenelli \etal\ \cite{Serenelli:2009yc}
(AGSS09 and AGSS09ph) based upon recent abundance estimates
\cite{Asplund:2009fu}, and assumed the AGSS09 model as our default.
We do the same here, performing a full numerical integration over the
radial profile of the Sun when determining the capture rates~\footnote{%
  Using the simpler Gould approximation~\cite{Gould:1992xx} would have
  yielded rates differing from the exact results by at most 6\%, as was
  shown in~\cite{EOSSnu}.
  }.

In modelling the dark matter halo, we assume a non-rotating isothermal
sphere with an rms speed of 270~km/s, a disk rotation speed of 220~km/s,
and a local dark matter density of 0.3~GeV/cm$^3$.
If the calculated neutralino relic density is below the WMAP observed
relic density, we assume that only a fraction of the local dark matter
density, equal to the ratio of the neutralino and WMAP dark matter
relic densities, is attributable to neutralinos.
As in~\cite{EOSSnu}, we do not address other halo models in this paper,
but note that our results would scale linearly with the local dark
matter density as equilibrium between capture and annihilation is
approached.

\begin{figure*}
  \insertdoublefig{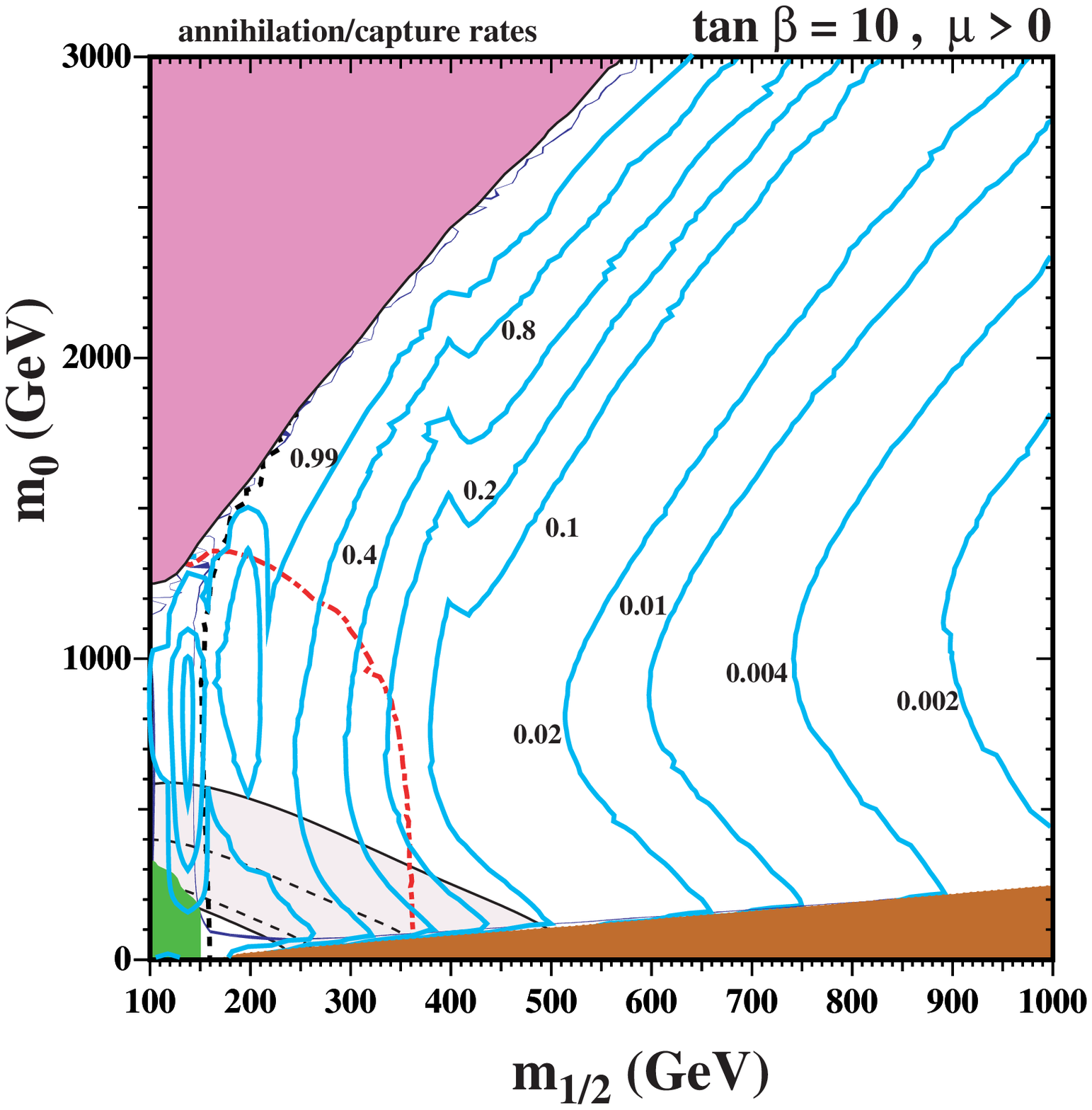}{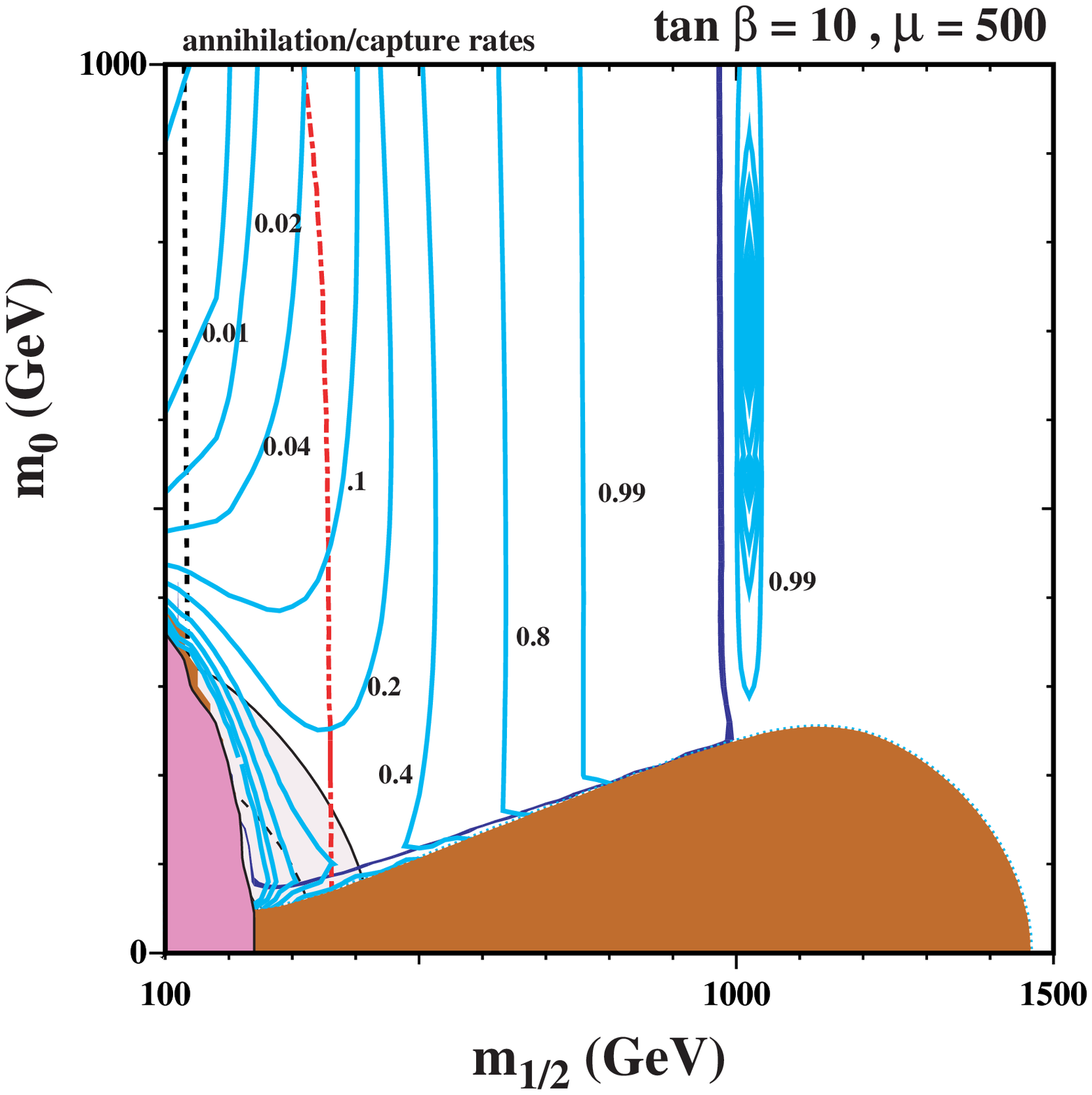}
  \caption{\it
    \coloronlinestatement
    The same CMSSM plane (left) and NUHM1 plane (right), displaying
    also contours of the ratio of solar dark matter annihilation and
    capture rates.
    Equilibrium corresponds to a ratio of unity, which is approached
    for small $m_{1/2}$ and large $m_0$ in the CMSSM, and near a
    vertical strip with $m_{1/2} \sim 1000$~GeV in the NUHM1 example.
    }
  \label{fig:capann}
\end{figure*}

However, as discussed extensively in~\cite{EOSSnu}, equilibrium is not
in general reached in the CMSSM, and an example is shown in the left
panel of \reffig{capann}, namely the $(m_{1/2}, m_0)$ plane for
$\tanb = 10, A_0 = 0$ and $\mu > 0$.
The dark blue contours are for different ratios of the annihilation and
capture rates, and we see that equilibrium is reached only for small
$m_{1/2}$ and large $m_0$. The right panel of \reffig{capann} shows the
corresponding contours in the NUHM1 $(m_{1/2}, m_0)$ plane for
$\tanb = 10, A_0 = 0$ and $\mu = 500$~GeV, where we see that equilibrium
is approached only near a vertical strip with  $m_{1/2} \sim 1000$~GeV.
In the following we calculate annihilation rates without relying on the
assumption of equilibrium.

The neutralino annihilations produce high-energy neutrinos which,
through interactions with matter, will induce muons in or around a
detector such as IceCube/DeepCore \cite{Ahrens:2003ix,Achterberg:2006md,
Resconi:2008fe,Abbasi:2009uz}.  The IceCube
detector has outfitted $\sim$1~km$^3$ of ice at the South Pole with
optical sensors to observe the Cerenkov light produced by the
passage of muons through the ice.  The large volume allows for
sensitivity to very low neutrino fluxes.  However, the relatively
large spacing between sensors severely limits the sensitivity to lower
energy muons (below 100~GeV) which are the largest
portion of the neutrino-induced muon spectra arising
from annihilations in the Sun.  To improve the sensitivity to these
important lower-energy muons, a portion of the IceCube volume,
referred to as DeepCore, has been outfitted with more densely packed
sensors.

The ability of IceCube/DeepCore to detect the flux from a given
supersymmetric model depends on various factors such as the muon
spectra, the neutrino backgrounds, and the method by which
IceCube/DeepCore will analyze their results.
However, to a very rough approximation, IceCube can
detect muon fluxes on the order of 10 or 10$^2$~/km$^2$/yr above
$\sim$100~GeV and DeepCore can detect muon fluxes
on the order of 10$^2$ or 10$^3$~/km$^2$/yr above $\sim$10~GeV.
In this paper, we primarily examine the latter case: total fluxes above
10~GeV.
DeepCore can detect muons down $\sim$10~GeV; however, the analysis
threshold may turn out to be somewhat higher:
Ref.~\cite{Barger:2011em} suggests an analysis threshold of
$\sim$35~GeV is reasonable, while Ref.~\cite{Danninger:2011pc} suggests
25--30~GeV is more likely (but possibly as low as 20~GeV) \footnote{%
  Triggering of two of IceCube/DeepCore's optical modules---which is
  possible for muons with energies as low as $\sim$10~GeV---is
  sufficient to detect a muon.  However, two-module events suffer from
  extremely poor angular resolution.  Muons that trigger three optical
  modules (which requires somewhat higher muon energies) yield much
  better track reconstruction and allow the analysis to be restricted
  to muons consistent with neutrinos coming from the direction of the
  Sun.
  }.
In addition, the efficiency of detecting muons (or, equivalently, the
effective area of the detector) falls with decreasing muon energy.
Our results are not significantly affected by our choice of a 10~GeV
threshold as we use only an order-of-magnitude estimate of the
IceCube/DeepCore sensitivity to the total muon flux above this energy.
Our results are only affected if the flux is predominantly just above
threshold (\eg\ between 10 and 25~GeV), which is only expected to be
the case for very light neutralinos ($m_\chi \ll 100$~GeV or
$m_{1/2} \ll 200$~GeV) \footnote{%
  The 10~GeV threshold was chosen and the bulk of the work done in this
  paper was performed prior to the availability of threshold estimates
  from Refs.~\cite{Barger:2011em,Danninger:2011pc}.
  As this choice does not significantly affect our results, we have
  chosen to keep the current threshold in our analysis.
  }.

We use the results of the WimpSim simulation \cite{wimpsim} (as used
within DarkSUSY~\cite{darksusy}) to calculate 
the spectra of the neutrinos produced by the annihilations and the
corresponding spectra of neutrino-induced muons in IceCube/DeepCore.
More details of the neutralino
capture/annihilation processes and the determination of the
neutrino/muon fluxes may be found in Ref.~\cite{EOSSnu}.

\section{\label{sec:NUHM1} Representative Studies in the NUHM1}

\subsection{Comparison with the CMSSM in the
            \texorpdfstring{$(m_{1/2}, m_0)$}{(m\_1/2,m\_0)} Plane}

\begin{figure*}
  \insertdoublefig{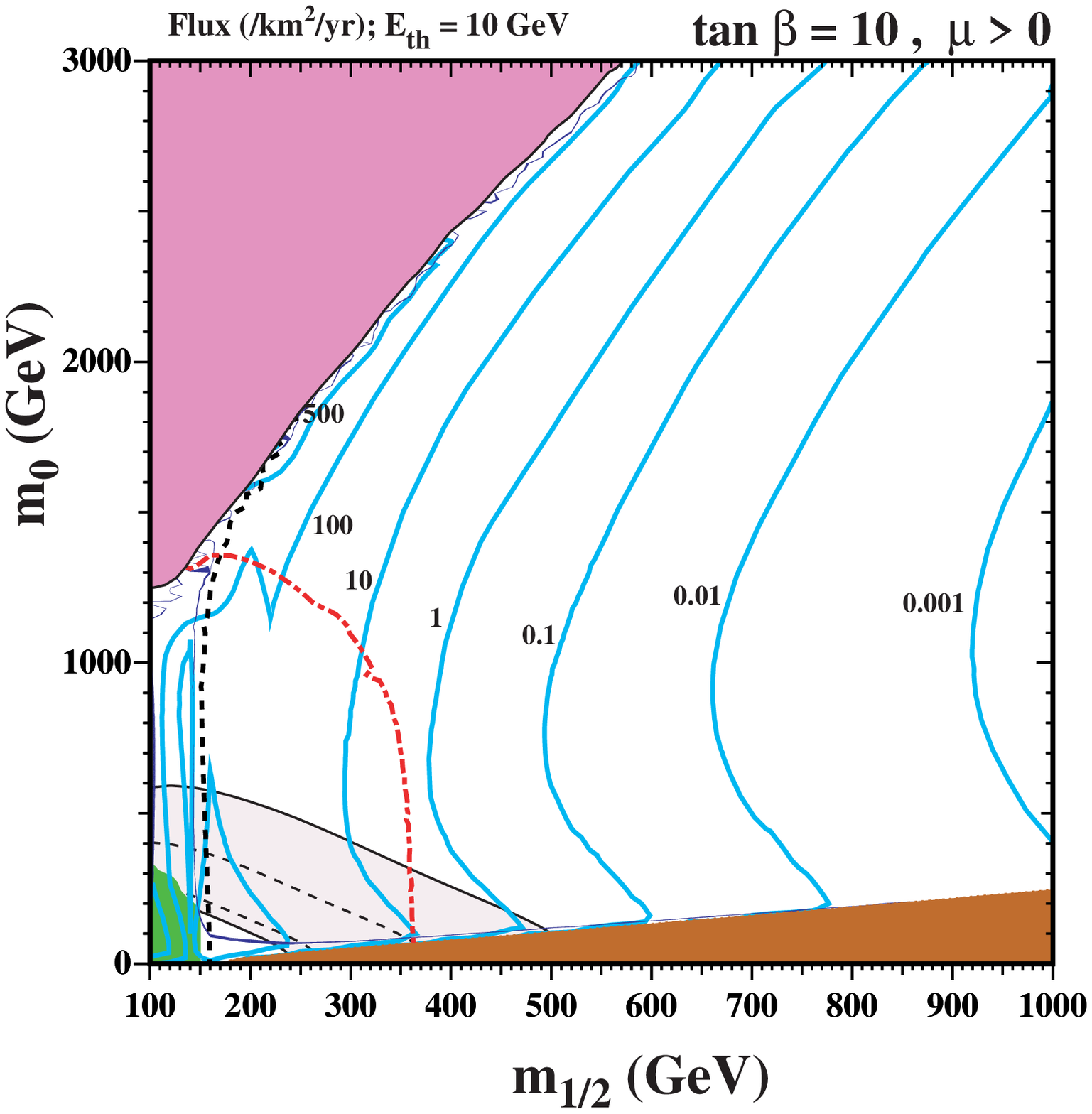}{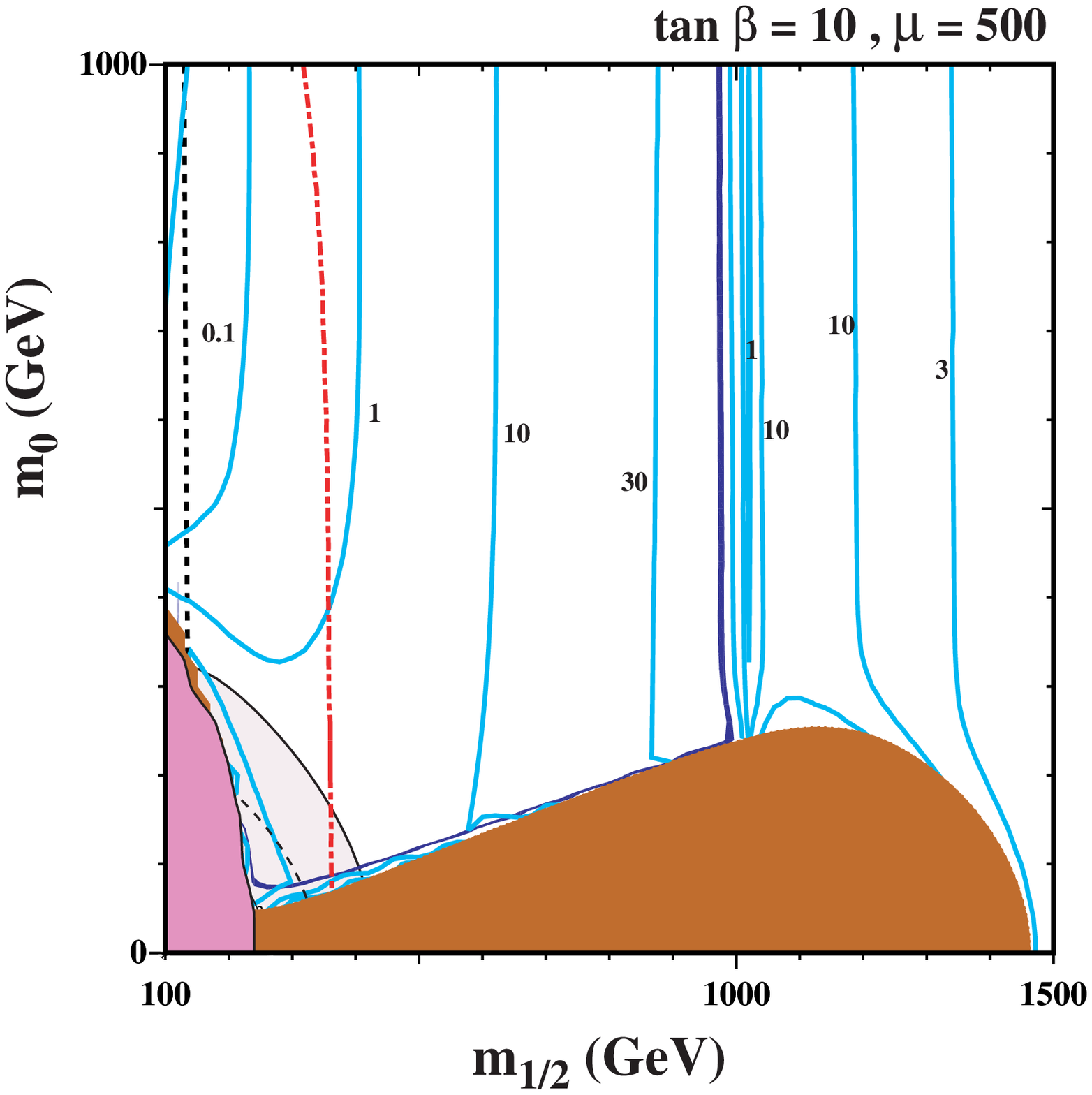}
  \caption{\it
    \coloronlinestatement
    The $(m_{1/2}, m_0)$ planes for $\tanb = 10$ and $A_0 = 0$ in
    (left) the CMSSM and (right) the NUHM1 for $\mu = 500$~GeV.
    The solid (light blue) lines are contours of the neutrino-induced
    muon fluxes above 10~GeV in units of events/km$^2$/yr.
    The shadings and other contours have the same meanings as in
    \reffig{capann}.
    }
  \label{fig:m12m0}
\end{figure*}

The left panel of \reffig{m12m0} displays the neutrino-induced muon
fluxes calculated in the $(m_{1/2}, m_0)$ plane for the typical CMSSM
scenario with $\tanb = 10, A_0 = 0$ and $\mu > 0$ introduced above. We
recall that this plane has two narrow strips where the LSP density falls
within the range allowed by WMAP and other
measurements~\cite{Komatsu:2010fb}, which are coloured turquoise. 
One is the coannihilation strip close to the boundary of the forbidden
charged-LSP region at low $m_0$, and the other is the focus-point strip
close to the EWSB boundary at large $m_0$. 

As was discussed extensively in~\cite{EOSSnu} and shown in the left
panel of Figs.~\ref{fig:sds} \&~\ref{fig:capann}, at large $m_{1/2}$ it
is usually not appropriate to assume that the LSP capture cross-section
is dominated by spin-dependent interactions, nor that there is
equilibrium between capture and annihilation. Accordingly, as already
mentioned, neither assumption is made in this and subsequent plots. The
solid (blue) lines are contours of the neutrino-induced muon fluxes
above 10~GeV in units of events/km$^2$/yr.
As seen in the left panel of \reffig{m12m0}, the neutrino rate is
potentially detectable in IceCube/DeepCore along a significant stretch
of the focus-point strip, but apparently undetectable along the portion
of the coannihilation strip that is compatible with the LEP Higgs
constraint. As discussed in~\cite{EOSSnu}, the neutrino fluxes in the
CMSSM are generally larger for $\tan \beta = 55$ along the
coannihilation strip (though still not observable in IceCube/DeepCore),
and smaller along the focus-point strip than they are for the
$\tan \beta = 10$ case shown. We also recall that when $\tan \beta =55$,
there is also a third region compatible with the dark matter density
constraint, namely a rapid-annihilation funnel at large $m_{1/2}$, where
the rate is far too small to be detectable by IceCube/DeepCore.

The right panel of \reffig{m12m0} displays the neutrino-induced muon
fluxes to be expected in a sample $(m_{1/2}, m_0)$ plane in the NUHM1,
again with $\tanb = 10$ and $A_0 = 0$, but now with fixed $\mu = 500$~GeV.
In this case, an excluded charged LSP region again appears at small $m_0$
though with a different shape from the CMSSM, and the EWSB boundary has
moved to small $m_{1/2}$ and $m_0$. Here it is $m_A^2$ that is driven
negative rather than $\mu^2$ as in the CMSSM. There is a region favoured
by $g_\mu - 2$ that is almost excluded by the LEP Higgs constraint, and
there is a coannihilation strip, as in the CMSSM, that follows the EWSB
boundary. There is also a near-vertical extension of the coannihilation
strip at $m_{1/2} \sim 1000$~GeV, that appears thanks to the freedom in
the NUHM1 of independently adjusting the $\mu$ parameter. It occurs
along a line where a particular relation between $\mu$ and $m_{1/2}$
induces level crossing in the neutralino mass matrix as $m_{1/2}$ is
increased relative to $\mu$, and thereby leads to increased
Higgsino-gaugino mixing that brings the LSP density down into the WMAP
range. To the right of this strip, the relic density falls below the
WMAP range. This strip is the only part of this particular NUHM1 plane
where the neutrino-induced muon flux approaches detectability in
IceCube/DeepCore, rising above 30~/km$^2$/yr, thanks to the enhanced
Higgsino-gaugino mixing.

One of the key questions in our analysis will be the extent to which the
freedom in the NUHM1,2 to vary $\mu$ and/or $m_A$  provides this and
other opportunities for IceCube/DeepCore detection of neutrinos that are
absent in the CMSSM.

\subsection{NUHM1 \texorpdfstring{$(\mu, m_{1/2})$}{(mu,m\_1/2)} Planes}

\begin{figure*}
  \insertdoublefig{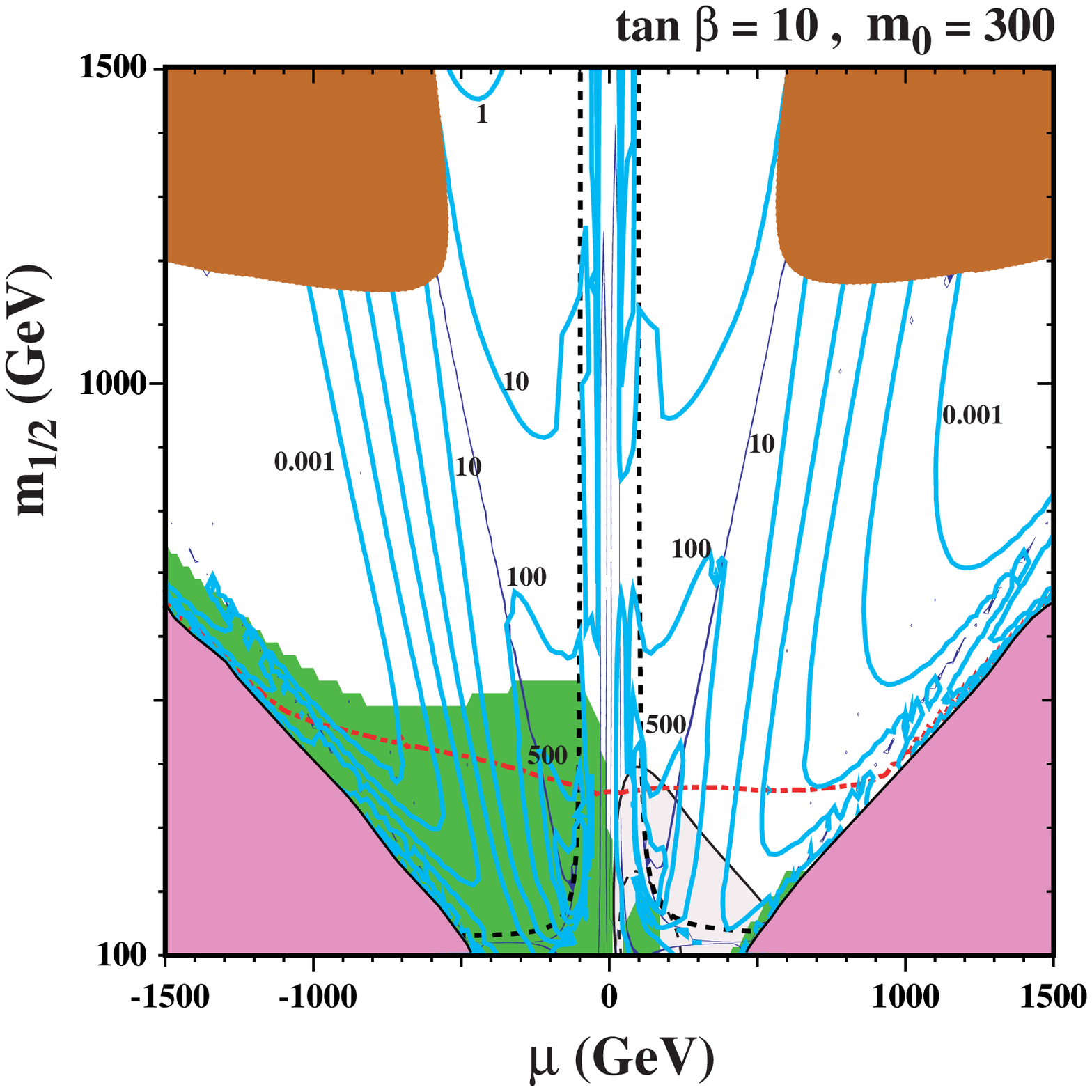}{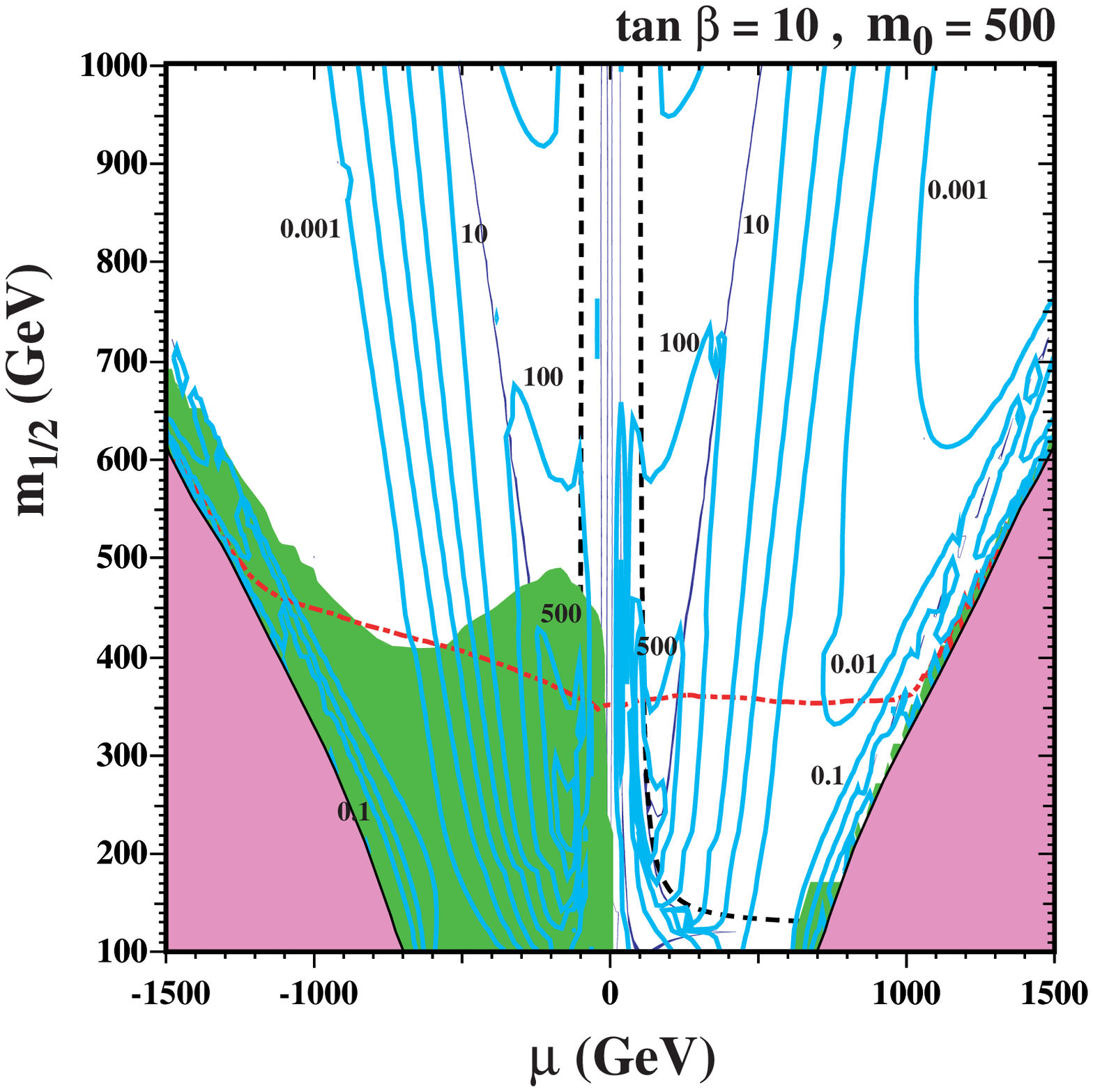}\\
  \insertdoublefig{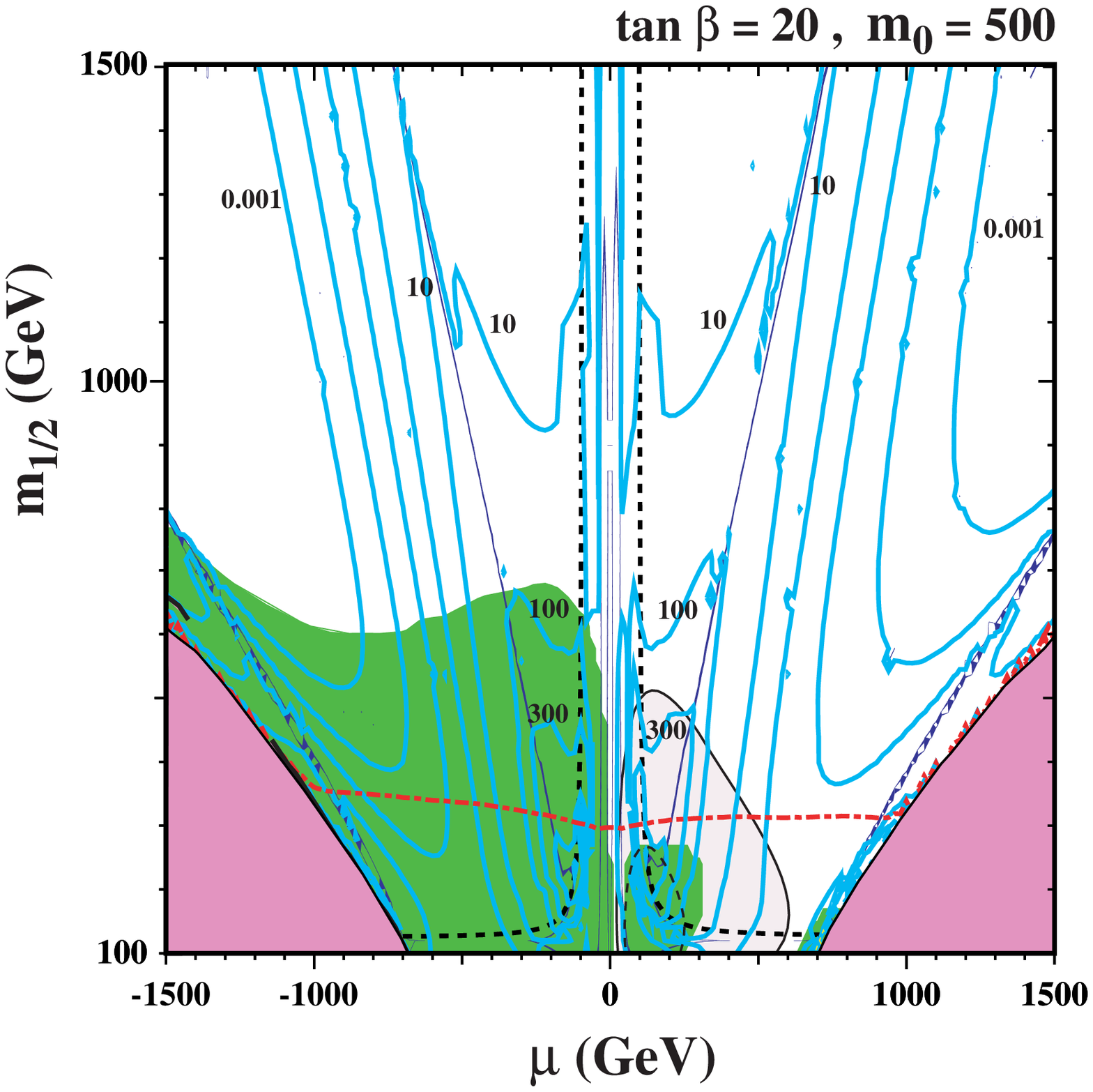}{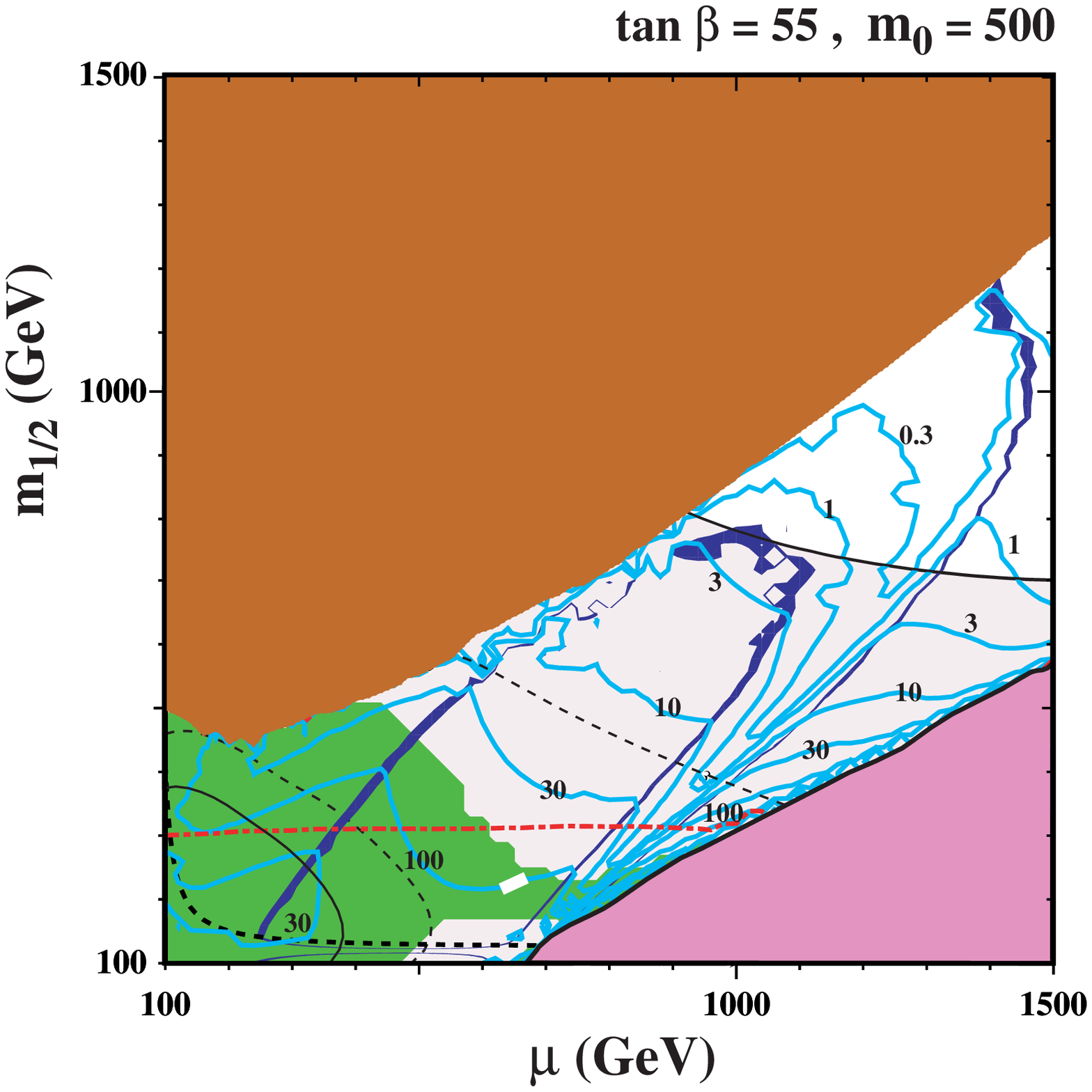}
  \caption{\it
    \coloronlinestatement
    The $(\mu, m_{1/2})$ planes in the NUHM1 for $A_0 = 0$ and
    (upper) $\tan \beta = 10$,
    (left) $m_0 = 300$~GeV and (right) $m_0 = 500$~GeV,
    (lower) $m_0 = 500$~GeV
    and (left) $\tan \beta = 20$ and (right) $\tan \beta = 55$.
    The shadings and contours have the same meanings as in
    \reffig{m12m0}.
    }
  \label{fig:mum12planes}
\end{figure*}

We now illustrate further the behaviour of this
IceCube/DeepCore-friendly WMAP strip with enhanced Higgsino-gaugino
mixing, first in some representative $(\mu, m_{1/2})$  planes.
\Reffig{mum12planes} displays planes for $\tanb = 10$ with (upper left)
$m_0 = 300$~GeV and (upper right) $m_0 = 500$~GeV: both values of $m_0$
are in the range favoured by a frequentist analysis of the NUHM1
parameter space \cite{mc3}.
In each case we see EWSB boundaries at large $|\mu|$ and small $m_{1/2}$
where $m_A^2 <0$, and regions with $\mu < 0$ that are disfavoured by
$b \to s \gamma$. In the upper left panel, we also see charged LSP
regions at large $|\mu|$ and $m_{1/2}$. In each case, there is a pair of
diagonal WMAP-compatible strips visible at $|\mu| \sim m_{1/2}/2$
where the Higgsino-gaugino mixing is enhanced, which are only weakly
dependent on $m_0$ and are compatible with the LEP Higgs constraint for
large enough $m_{1/2}$. The neutrino-induced muon flux above 10~GeV
is potentially detectable along essentially all of these diagonal
WMAP-compatible strips in both panels, though decreasing as $|\mu|$ and
$m_{1/2}$ increase.
These strips constitute extensions of the IceCube/DeepCore-friendly
strip seen in the right panel of \reffig{m12m0}, which has
$\mu = 500$~GeV and $m_{1/2} \sim 1000$~GeV, to different values of
these NUHM1 parameters. The red dot-dashed curve is the contour for
$m_h = 114$~GeV and one should preferably lie above this curve, though
one should recall that there is a 1.5~GeV uncertainty in the theoretical
calculation of $m_h$.  Nevertheless, fluxes above the Higgs limit still
reach above 500~/km$^2$/yr.
For $m_0 = 300$~GeV, the $g-2$ constraint would prefer lower values of
$m_{1/2}$ in potential conflict with the Higgs bound -- though the muon
fluxes are quite large where both constraints are satisfied.
At $m_0 = 500$~GeV, the $g-2$ is not satisfied within 2$\sigma$ anywhere
on the plot. We note that there is another WMAP strip slightly above the
region with no EWSB. This strip corresponds to the heavy Higgs funnel
where $2 m_\chi \approx m_A$ and would not be present in the CMSSM at
$\tan \beta = 10$. However, neutrino-induced muon fluxes are too small
along these strips to be observed in IceCube/DeepCore. 

Similar diagonal strips with enhanced Higgsino-gaugino mixing are seen
in the lower left panel of \reffig{mum12planes} for $\tanb = 20$ and
$m_0 = 500$~GeV, though the muon fluxes are generally smaller. However,
the diagonal strip for $\mu > 0$ has a part where $g_\mu - 2$ lies
within the favoured range and the muon flux is $> 300$~/km$^2$/yr. As in
the previous panels of \reffig{mum12planes}, the WMAP-compatible strips
corresponding to the funnel and close to the EWSB boundary have muon
fluxes which are unobservably low. In the lower right panel of
\reffig{mum12planes} for $\tanb = 55$ and $m_0 = 500$~GeV we see that,
whilst the EWSB boundary is located similarly to the previous panels,
the charged LSP region has advanced considerably, as has the region
excluded by $b \to s \gamma$. At this value of $\tan \beta$, $\mu < 0$
is not consistent with RGE running.
We also note that the diagonal strip with enhanced Higgsino-gaugino
mixing has merged with the rapid-annihilation funnel, with both sides
of the funnel now easily discernible. 
The muon fluxes where $b \to s \gamma$ is acceptable are
$\lae 50$~/km$^2$/yr and large enough to be detectable by
IceCube/DeepCore in the lower part of the transition strip, for values
of $m_{1/2}$  above the bounds imposed by the LEP Higgs constraints and
well with the 1$\sigma$ bounds from $g-2$ (denoted by the dashed black
curves). The entire region with $m_{1/2}$ below 700-800~GeV yields
values of $g_\mu - 2$ in the range favoured by experiment.
In contrast to the cases with lower $\tan \beta$, the funnel region now
also has neutrino fluxes that are sufficiently large to be
observable~\footnote{%
  We note that equilibrium is established along the transition strip in
  all cases, but not along the funnel except at very large $\tan \beta$.
  We also note that spin-dependent scattering is dominant at low $|\mu|$
  for $\tanb = 10$ and 20, but subdominant at larger $|\mu|$ for all
  values of $\tanb$.
  }.

\subsection{NUHM1 \texorpdfstring{$(\mu, m_0)$}{(mu,m\_0)} Planes}

\begin{figure*}
  \insertdoublefig{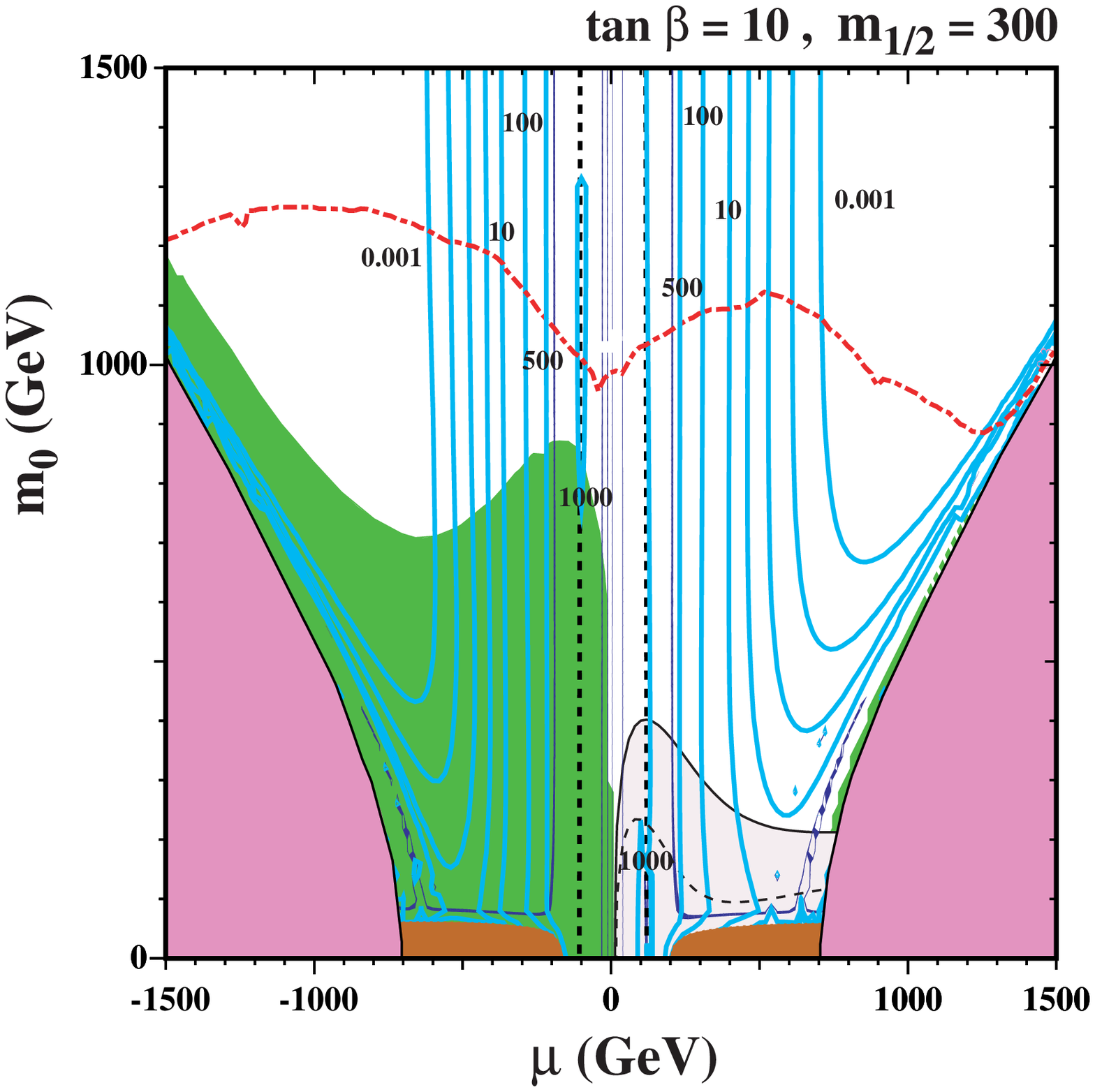}{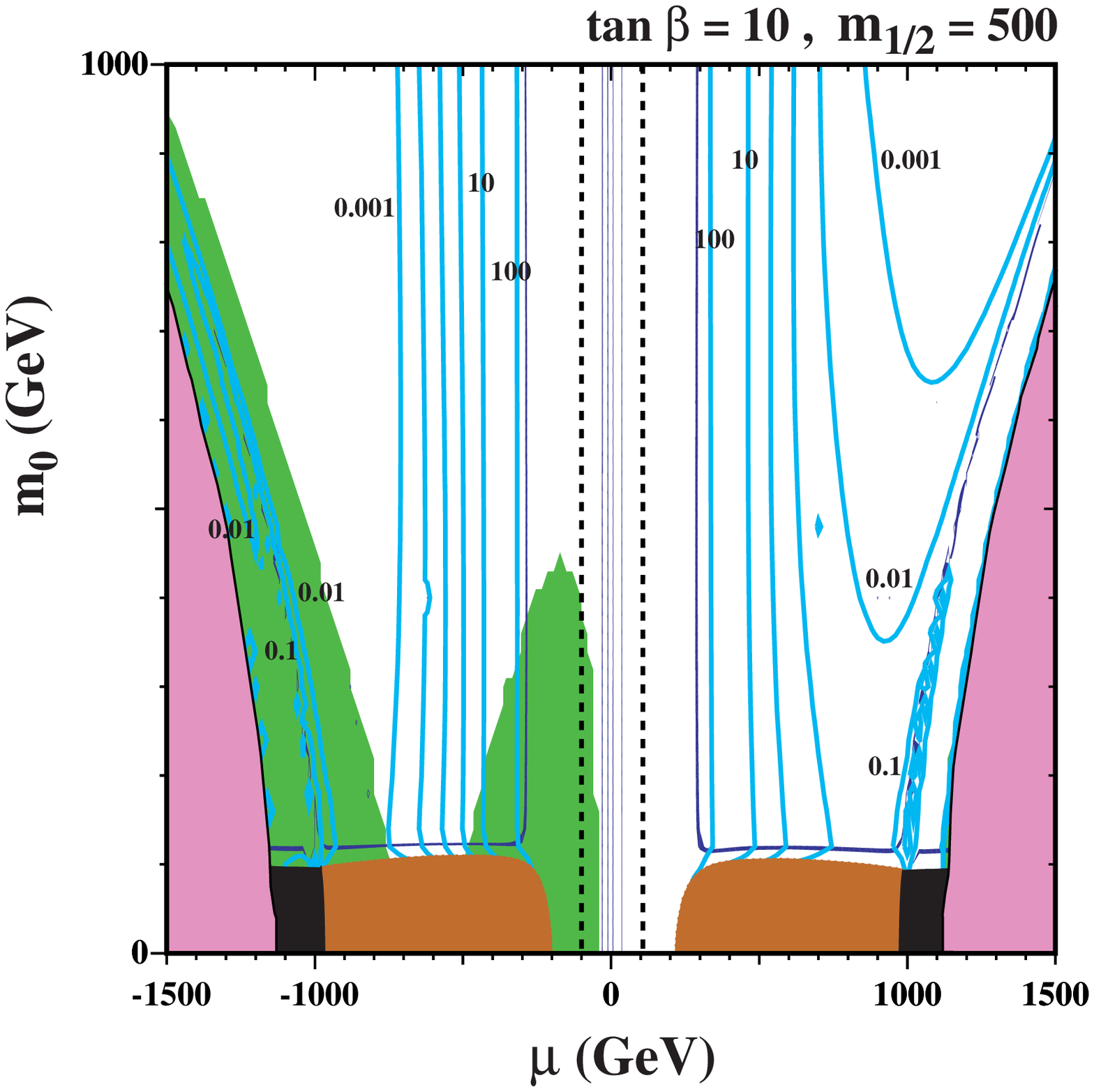}\\
  \insertdoublefig{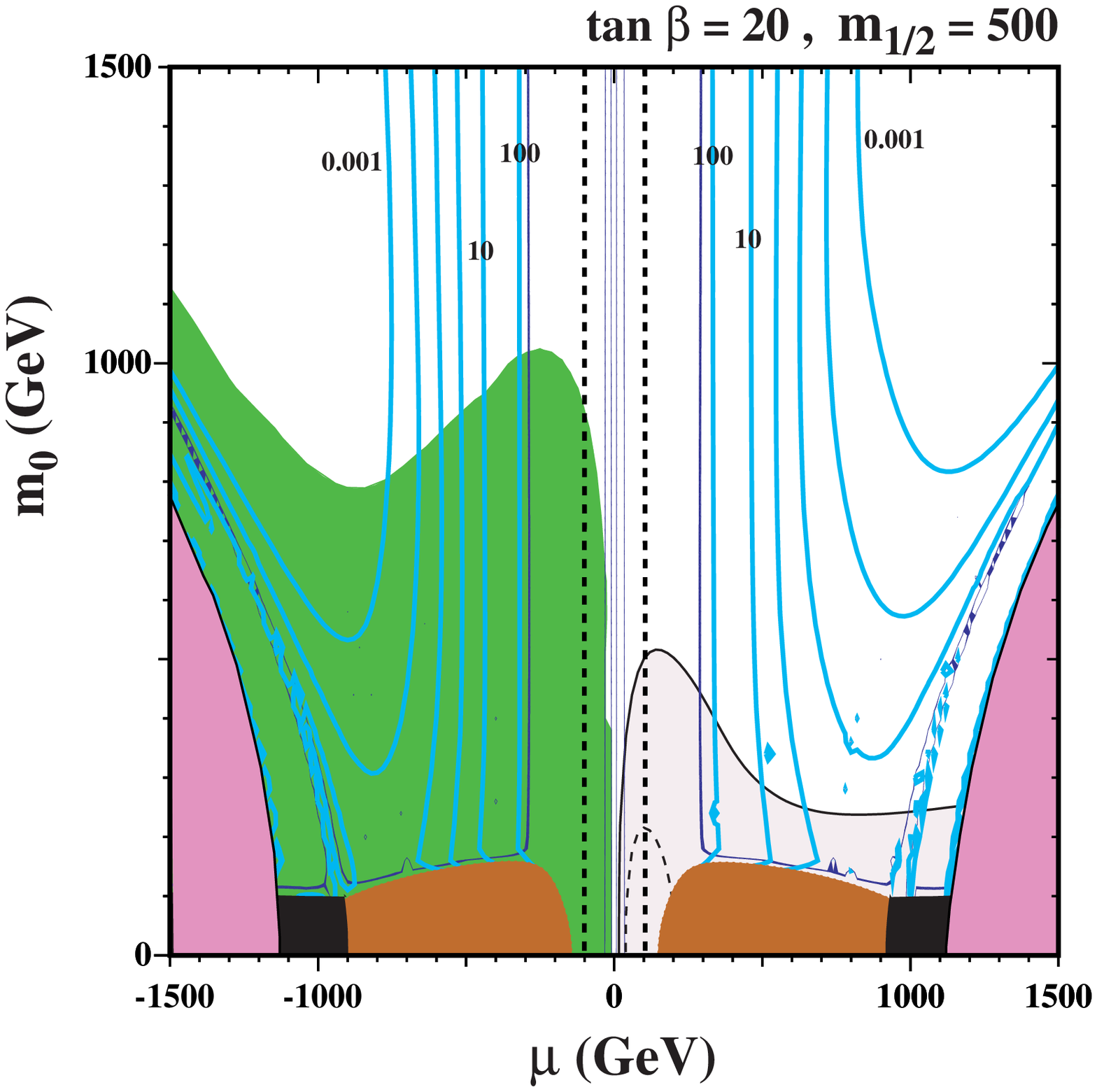}{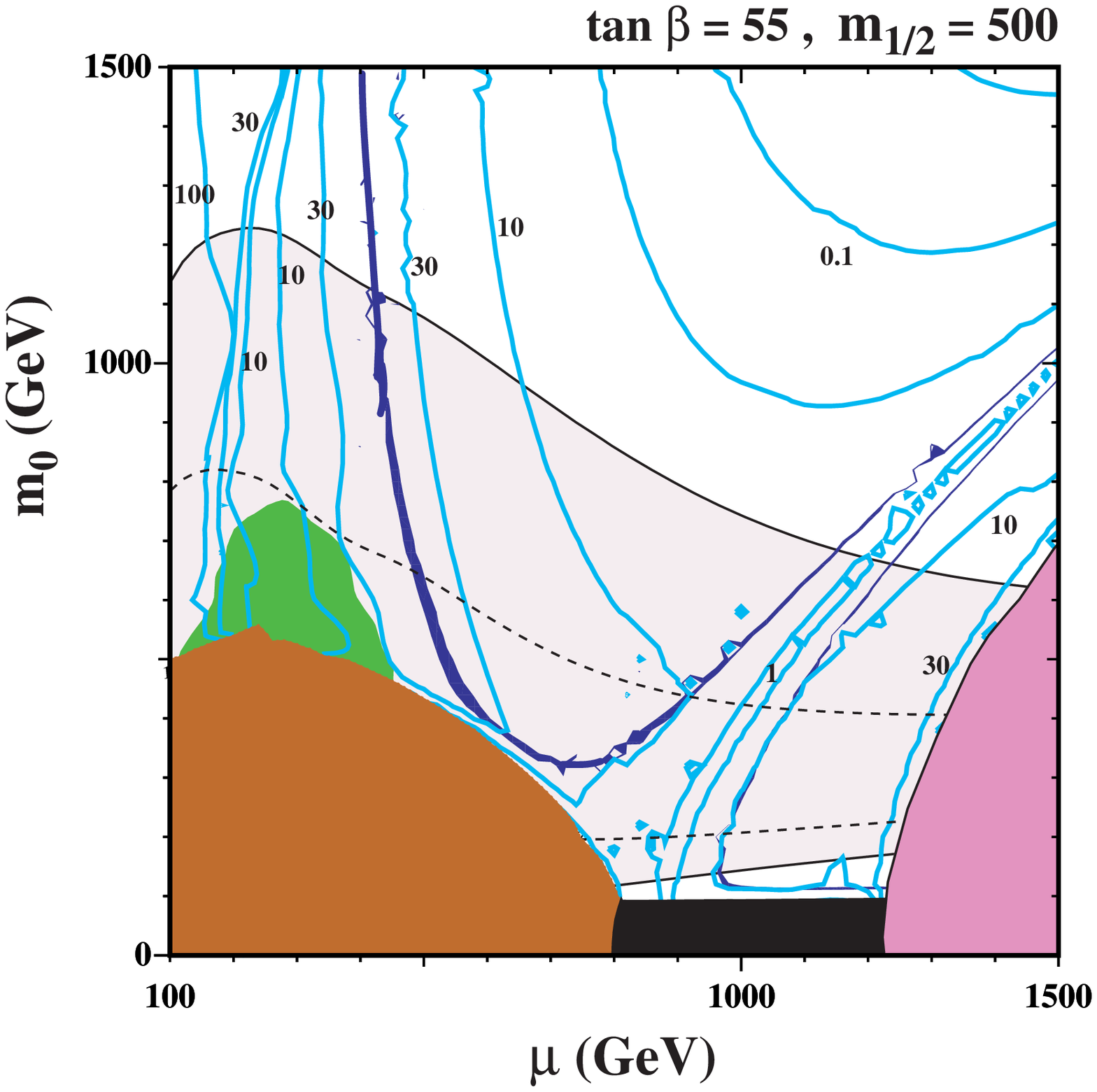}
  \caption{\it
    \coloronlinestatement
    The $(\mu, m_0)$ planes in the NUHM1 for $A_0 = 0$ and
    (upper) $\tan \beta = 10$,
    (left) $m_{1/2} = 300$~GeV and (right) $m_{1/2} = 500$~GeV,
    (lower) $m_{1/2} = 500$~GeV
    and (left) $\tan \beta = 20$ and (right) $\tan \beta = 55$.
    In the black regions the LSP is the right-handed selectron or smuon, 
    and the other shadings and contours have the same meanings as in
    \reffig{m12m0}.
    }
  \label{fig:mum0planes}
\end{figure*}

In order to explore further the IceCube/DeepCore-friendly region, in
\reffig{mum0planes} we display NUHM1 planes for $\tanb = 10$ with
(upper left) $m_{1/2} = 300$~GeV and (upper right) $m_{1/2} = 500$~GeV
(again, both values of $m_{1/2}$ are in the range favoured by a
frequentist analysis of the NUHM1 parameter space \cite{mc3}), and for
$m_{1/2} = 500$~GeV with (lower left) $\tanb = 20$ and (lower right)
$\tanb = 55$. We see in both the upper panels EWSB boundaries at large
$|\mu|$ and small $m_0$, charged LSP regions at smaller $|\mu|$ and
$m_0$, and regions excluded by $b \to s \gamma$ when $\mu < 0$.
When $m_{1/2} = 500$~GeV, there is a small region (shaded black)
between the stau LSP region and the EWSB boundary where the LSP is
a right-handed selectron (or smuon). 
As in the right panel of \reffig{m12m0}, we see in both of the upper
panels WMAP-compatible strips at roughly fixed values of $|\mu|$ related
to the values of $m_{1/2}$, along which the muon flux above 10~GeV is
potentially detectable: above 500 events/km$^2$/yr for (upper left)
$m_{1/2} = 300$~GeV and $\mu \sim 200$~GeV, and above
100 events/km$^2$/yr for (upper right) $m_{1/2} = 500$~GeV and
$\mu \sim 300$~GeV.
As already commented, along these strips the fluxes are enhanced 
by Higgsino-gaugino mixing. We note that only part of the strip for
$m_{1/2} = 300$~GeV is compatible with the LEP Higgs constraint,
whereas all the $m_{1/2} = 500$~GeV strip is compatible. There are no
parts of the IceCube/DeepCore-friendly strips in regions favoured by
$g_\mu - 2$. In both panels, there are WMAP-compatible extensions of
these strips to larger $\mu$ at $m_0 \sim 100$~GeV, due to
coannihilation, which segue into funnel strips close to the EWSB
boundary. However, these do not yield neutrino fluxes interesting for
IceCube/DeepCore. In both upper panels, equilibrium is established and
spin-dependent scattering is dominant along the transition strip, whilst
the opposite is true along the funnel.

Turning now to the lower panels in \reffig{mum0planes}, we see that as
$\tanb$ increases with fixed $m_{1/2} = 500$~GeV the EWSB boundary moves
away to larger $|\mu|$, the charged LSP region rises to larger $m_0$ and
(for $\tanb = 55$) the $b \to s \gamma$ exclusion extends to $\mu > 0$
as well as the visible parts of the half-plane with $\mu < 0$.
In the lower left panel for $m_{1/2} = 500$~GeV and $\tanb = 20$ we see
near-vertical WMAP-compatible strips with $|\mu| \sim 300$~GeV where 
Higgsino-gaugino mixing is enhanced and the neutrino flux is favourable
for IceCube/DeepCore, and part of the strip for $\mu > 0$ is also
favoured by $g_\mu - 2$. This interesting region then bends into a
near-horizontal coannihilation strip, resembling those in the upper
panels, where the neutrino-induced muon flux is mostly unfavourable. 
As $|\mu|$ increases at low $m_0$, the LSP changes from a mostly
right-handed stau to a right-handed $\tilde{e}/\tilde{\mu}$. The muon
flux is also small in the funnel that follows the stau LSP boundary.
Turning finally to the lower right panel in \reffig{mum0planes}, we
clearly see a two-sided diagonal funnel where the rapid annihilation via
direct-channel heavy Higgs poles brings the relic density into the
WMAP-compatible range, albeit with a relatively low muon flux. 
We also see that the strip with enhanced Higgsino-gaugino mixing is less
vertical and with a lower flux than previously, though still
IceCube/DeepCore-friendly. In this case, the $g-2$ constraint is
satisfied over much of the plane.

\subsection{NUHM1 \texorpdfstring{$(m_A, m_{1/2})$}{(m\_A,m\_1/2)} Planes}

\begin{figure*}
  \insertdoublefig{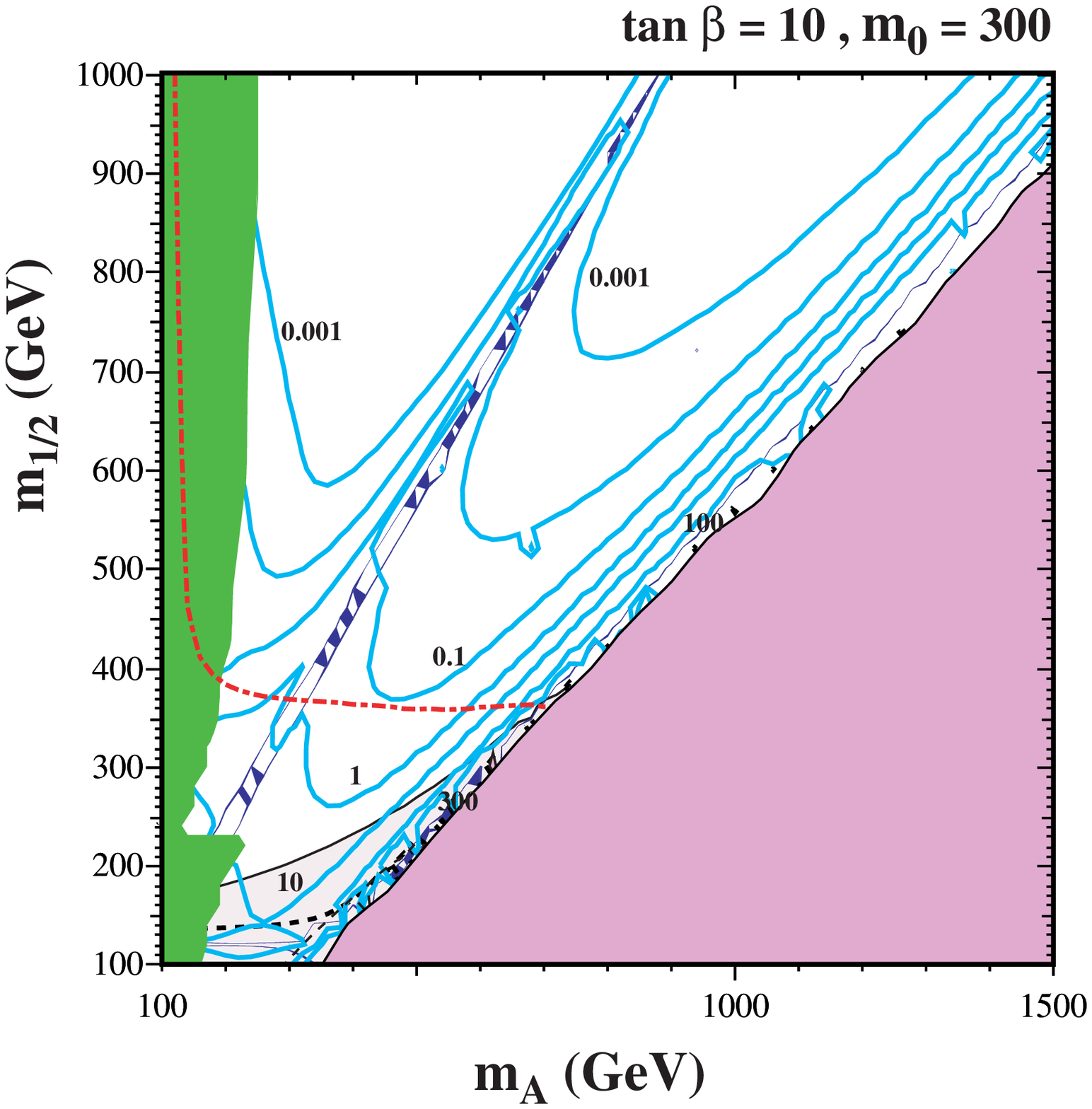}{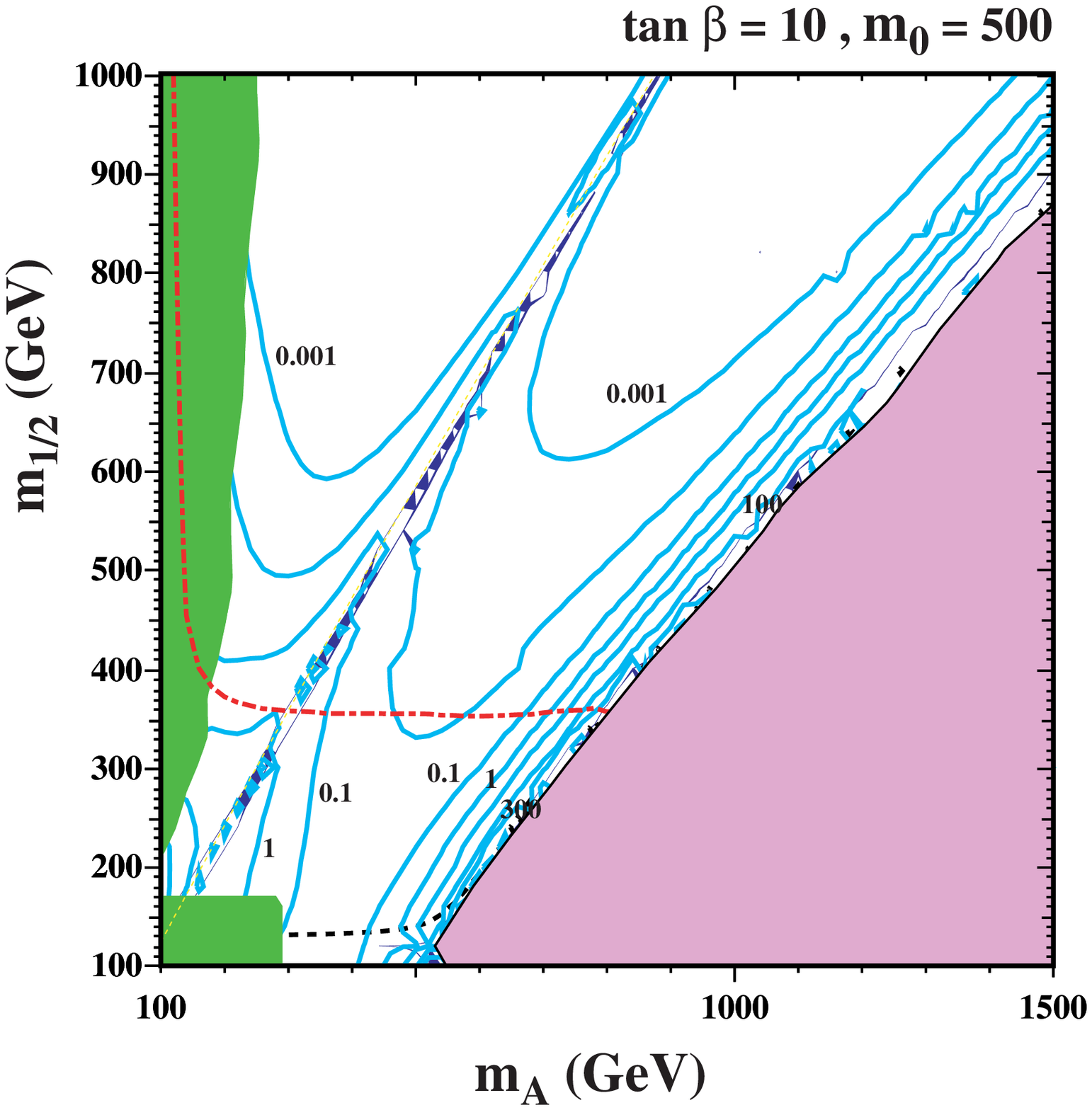}\\
  \insertdoublefig{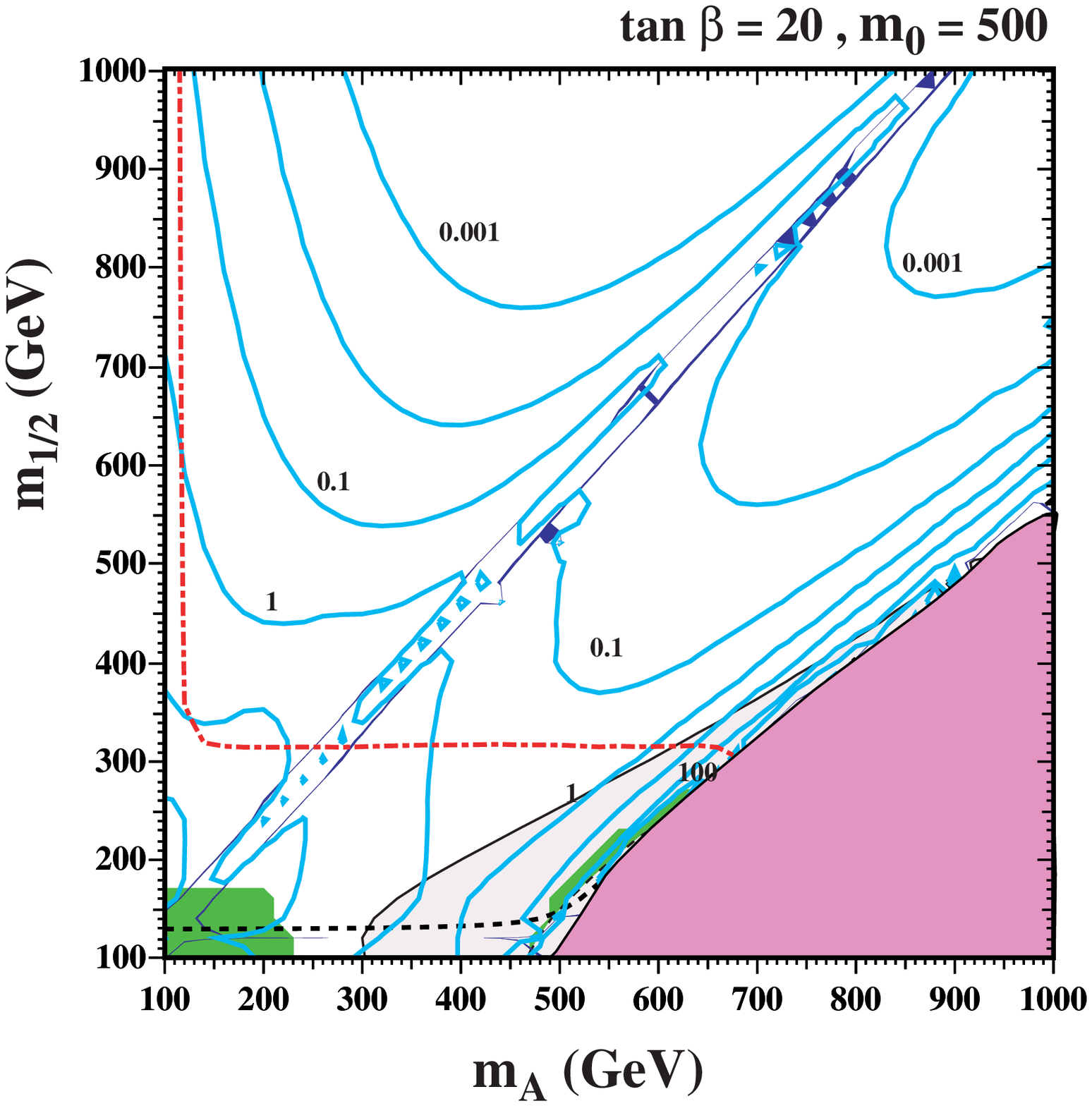}{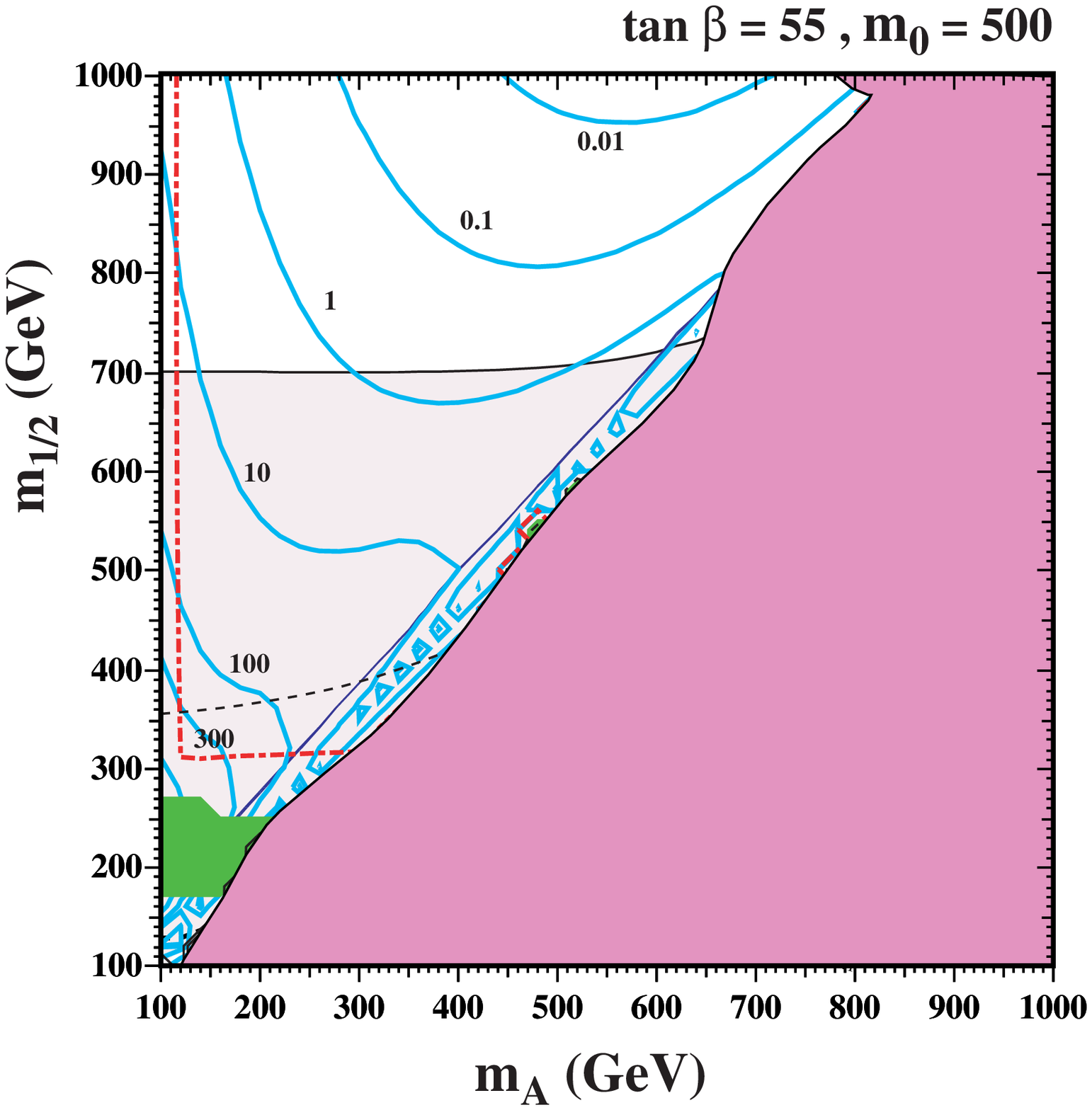}
  \caption{\it
    \coloronlinestatement
    The $(m_A, m_{1/2})$ planes in the NUHM1 for $A_0 = 0$ and
    (upper row) $\tanb = 10$ and (left) $m_0 = 300$~GeV,
    (right) $m_0 = 500$~GeV,
    (lower row) for $m_0 = 500$~GeV and (left) $\tanb = 20$
    and (right) $\tanb = 55$.
    The shadings and contours have the same meanings as in
    \reffig{m12m0}.
  }
  \label{fig:mAm12planes}
\end{figure*}

\Reffig{mAm12planes} displays some representative $(m_A, m_{1/2})$
planes in the NUHM1 for $A_0 = 0$ and $\tanb = 10$ with (upper left)
$m_0 = 300$~GeV and (upper right) $m_0 = 500$~GeV. 
In both cases, we see that the EWSB requirement excludes a triangular
region at large $m_A$ and small $m_{1/2}$ where $\mu^2$ is driven
negative, whereas $b \to s \gamma$ excludes a band at small $m_A$.
There is a region favoured by $g_\mu - 2$ in the left plane that is
again almost excluded by the LEP Higgs constraint.
In each case, a WMAP-compatible strip runs parallel to the EWSB
boundary, extending to small $m_A$ at $m_{1/2} \sim 125$~GeV.
There is also a diagonal funnel where LSPs annihilate rapidly though
direct-channel Higgs poles, because $m_\chi \sim m_A/2$, and there are
WMAP strips on either side of this funnel. The neutrino flux is very
unfavourable along the funnel, since the neutrino-induced muon flux 
above 10~GeV is $\ll 10$ events/km$^2$/yr. However, a signal may be 
observable in IceCube/DeepCore along the EWSB boundary, where the muon
flux above 10~GeV  lies in the range 10 to 100 events/km$^2$/yr.
As we have seen before,  equilibrium is established and spin-dependent
scattering is dominant along the transition strip near the EWSB
boundary, in contrast to the funnel regions where equilibrium is not
established and spin-independent scattering is dominant. 
 
Turning now to the lower row of plots in \reffig{mAm12planes} for (left)
$\tanb = 20$ and (right) $\tanb = 55$, both with $m_0 = 500$~GeV, we see
that the EWSB boundary is broadly similar, whereas the region forbidden
by $b \to s \gamma$ is much reduced. The WMAP-compatible
rapid-annihilation funnel is clearly visible for $\tanb = 20$, but folds
into the EWSB boundary strip for $\tanb = 55$. As in the previous cases,
the neutrino flux is unobservably small along the rapid-annihilation
funnel, but may be observable along the EWSB boundary strip for
$\tanb = 20$. We note that a portion of the IceCube/DeepCore-friendly
region for $\tanb = 20$ is favoured by $g_\mu - 2$.
For $\tan \beta = 55$, muon fluxes are $\gae 10$~/km$^2$/yr for
$m_{1/2} \lae 500$~GeV along the transition/funnel and compatible with
$g-2$. We note that this case is also interesting because in much of the
allowable plane capture is dominated by spin-independent scattering.

\subsection{NUHM1 \texorpdfstring{$(m_A, m_0)$}{(m\_A,m\_0)} Planes}

\begin{figure*}
  \insertdoublefig{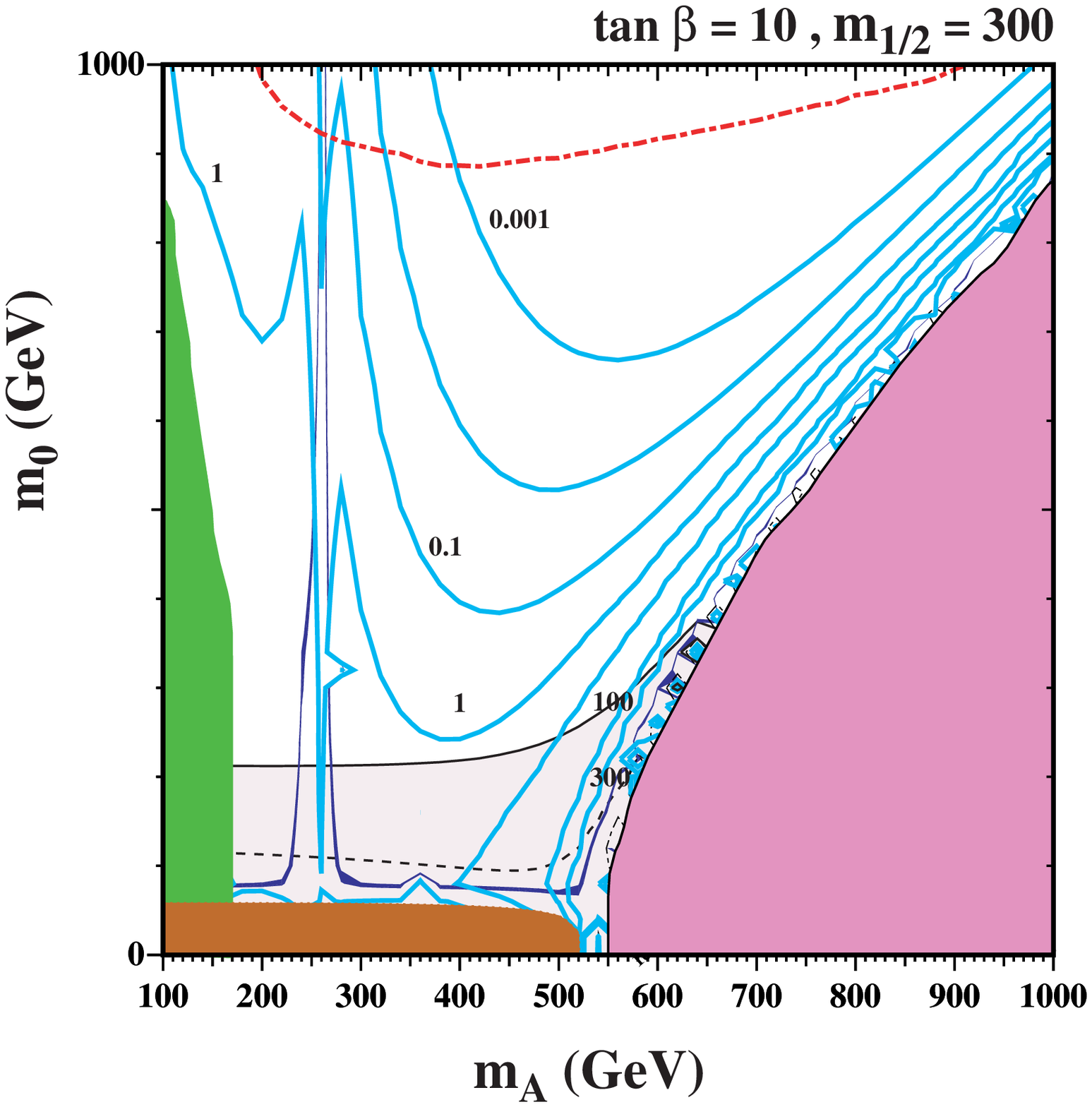}{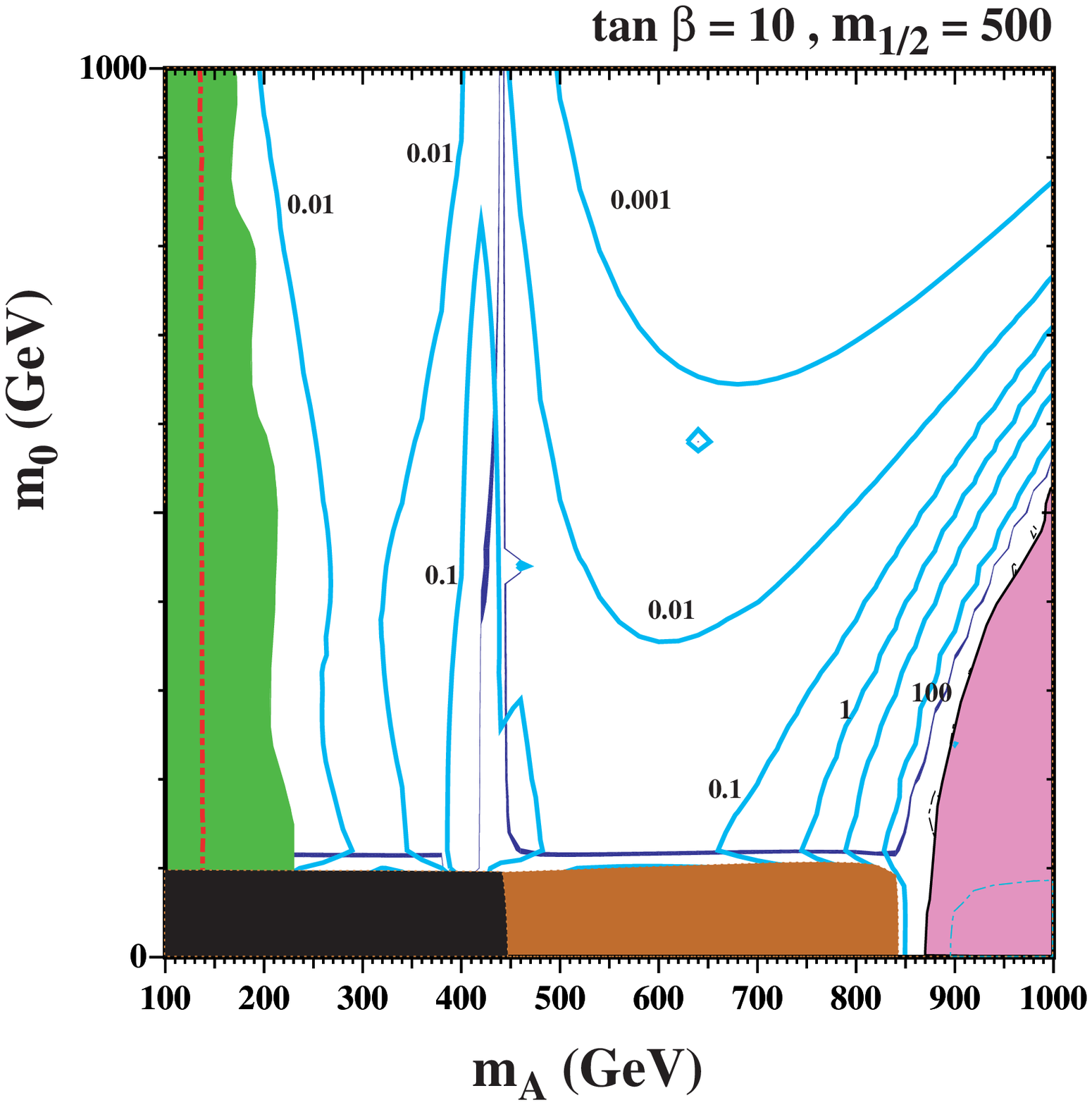}\\
  \insertdoublefig{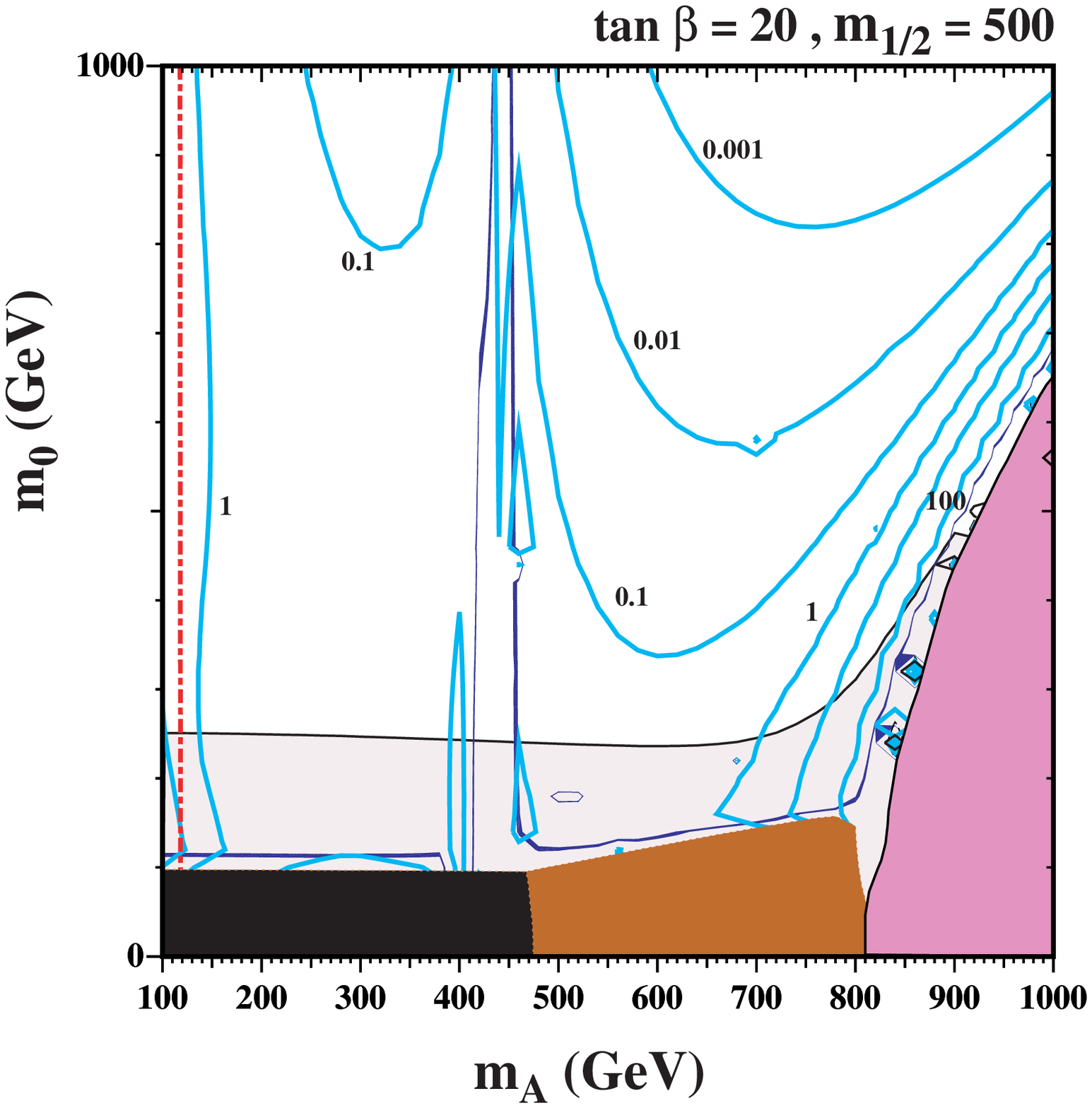}{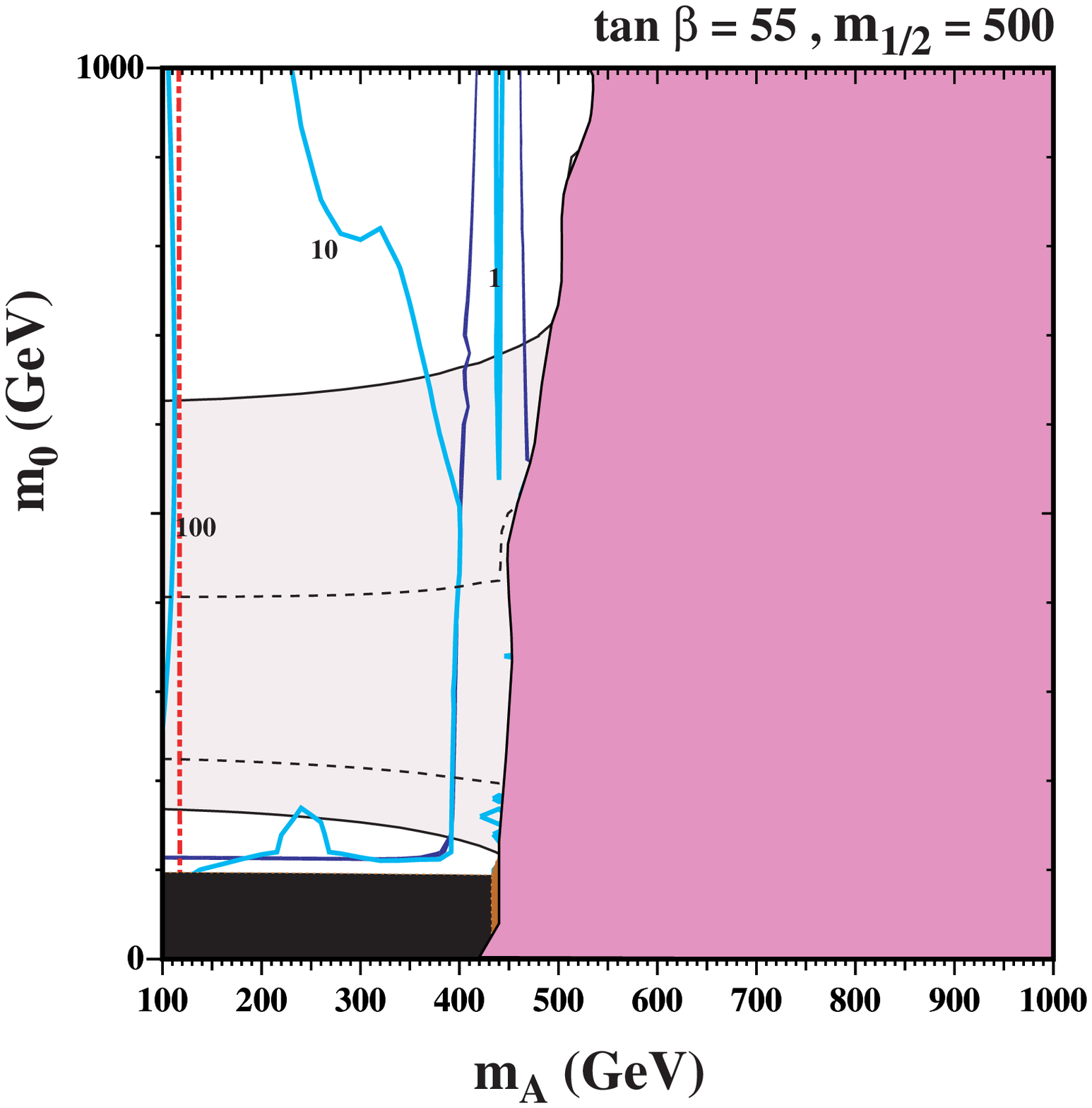}
  \caption{\it
    \coloronlinestatement
    The $(m_A, m_0)$ planes in the NUHM1 for $A_0 = 0$ and
    (upper row) $\tanb = 10$ and (left) $m_{1/2} = 300$~GeV,
    (right) $m_{1/2} = 500$~GeV,
    and (lower row) for $m_{1/2} = 500$~GeV and
    (left) $\tan \beta = 20$, (right) $\tan \beta = 55$.
    The shadings and contours have the same meanings as in
    \reffig{mum0planes}.
    }
  \label{fig:mAm0planes}
\end{figure*}

\Reffig{mAm0planes} shows some sample NUHM1 $(m_A, m_0)$ planes with
fixed $\tanb$ that exhibit rapid-annihilation funnels. In these cases,
the funnels appear as essentially vertical double strips on either side
of the line where $m_A = 2 m_\chi$. In each case, we also note the
presences of WMAP-compatible strips close to the EWSB boundary where
$\mu^2 = 0$. The two are attached by coannihilation strips at
$m_0 \sim 100$~GeV. When $m_{1/2} = 500$~GeV, coannihilation to the left
of the funnel at low $m_A$ is dominated by selectrons/smuons which are
the LSPs in the lower left corners. 
In the upper left panel for $\tanb = 10$ and $m_{1/2} = 300$~GeV, we
also see a region at low $m_0$ that is favoured by $g_\mu -2$, but this
is in a region disfavoured by the LEP Higgs limit. In this plane, the
only WMAP-compatible region allowed by the other constraints is up the
funnel at large $m_0$, where the neutrino flux in unobservably low.
In the upper right panel for $m_{1/2} = 500$~GeV, the Higgs constraint
is irrelevant, but the neutrino flux is still very low in all the
WMAP-compatible region except along the strip close to the EWSB boundary.
The same is true in the lower left plot for $\tanb = 20$ and
$m_{1/2} = 500$~GeV, where we note that a part of this strip is inside
the region favoured by $g_\mu - 2$.
However, as shown in the lower right plot, when $\tanb$ is increased to
55 with the same value of $m_{1/2}$, the EWSB boundary moves so close to
the rapid-annihilation funnel that there is no IceCube/DeepCore-friendly
region.

\subsection{NUHM1 Summary}

We have found in the above analysis that interesting neutrino fluxes
may arise from enhanced Higgsino-gaugino mixing. This possibility arose
in the CMSSM in the focus-point region along the EWSB boundary, but may
arise in regions of the NUHM1 parameter space that are far from this
boundary, corresponding in general to larger values of $m_{1/2}$.
The EWSB boundary region may also yield interesting fluxes when
$\mu^2 \to 0$ at the boundary. One of the other mechanisms present in
the NUHM1 for bring the relic LSP density into the WMAP-compatible range
is rapid annihilation through direct-channel Higgs poles. However, we
have found that this region generally yields low neutrino fluxes.

\section{\label{sec:NUHM2} Representative Studies in the NUHM2}

\begin{figure*}
  \insertdoublefig{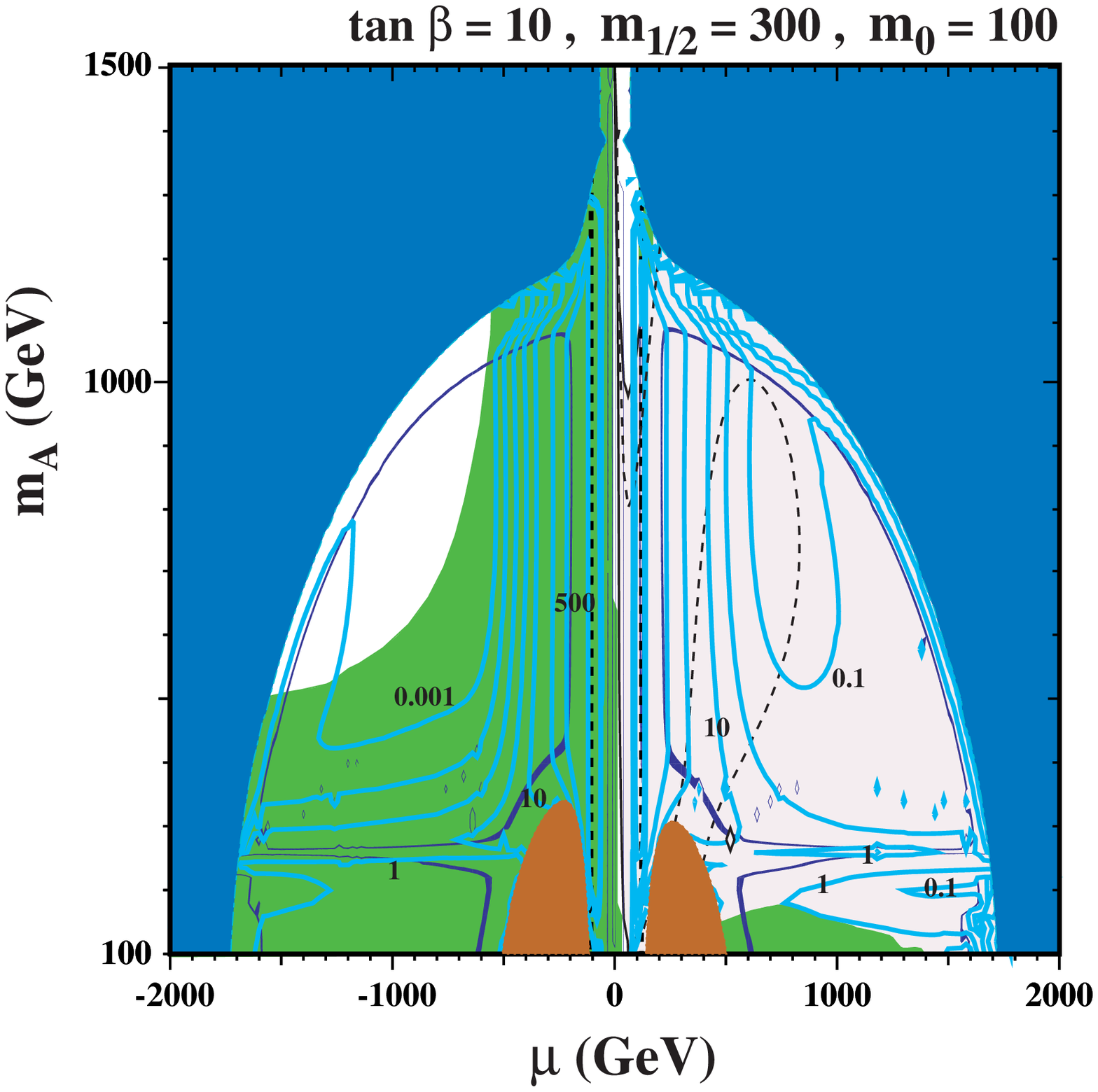}{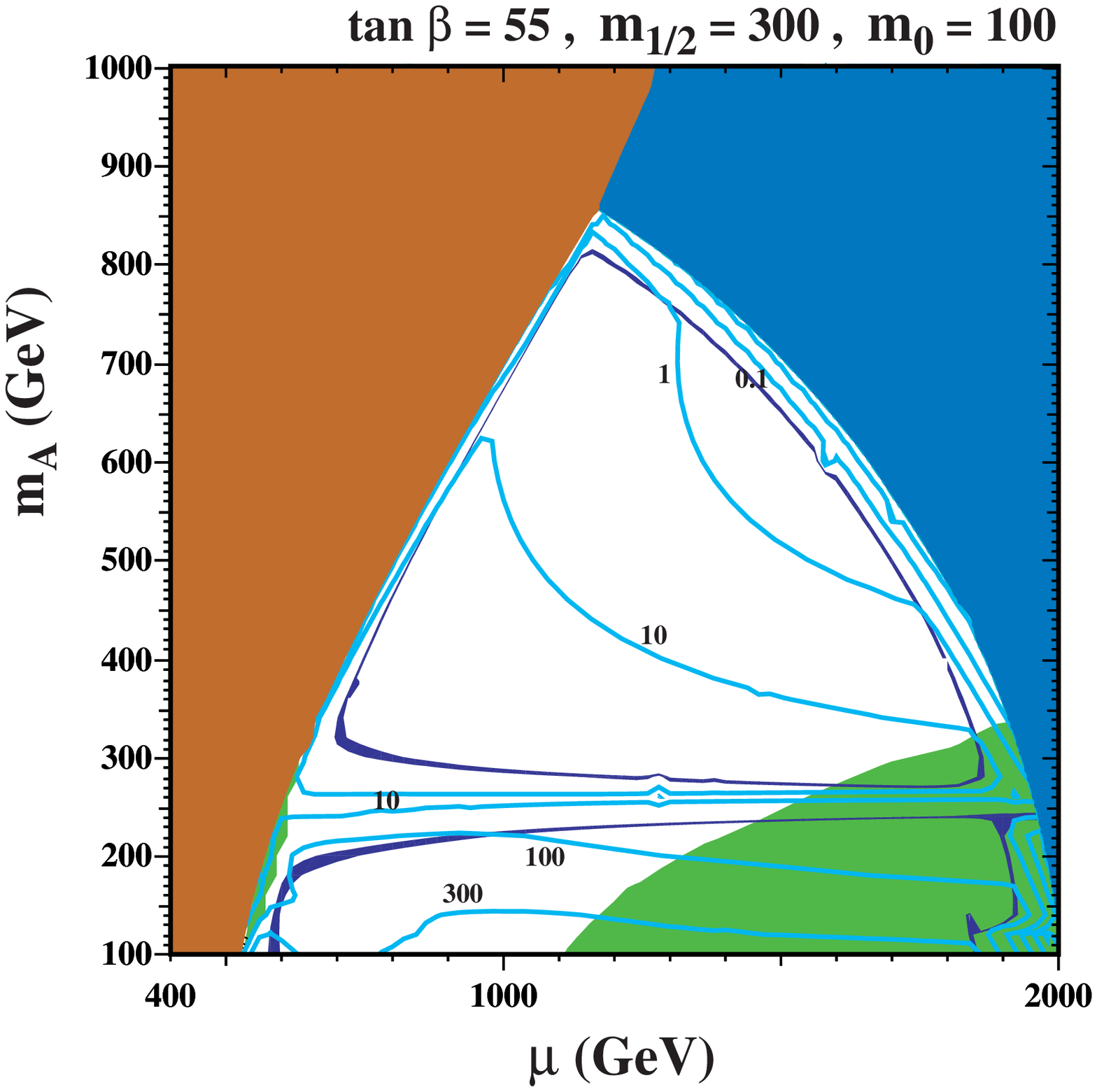}
  \caption{\it
    \coloronlinestatement
    The $(\mu, m_A)$ planes in the NUHM2 for $A_0 = 0, m_{1/2} = 300$~GeV
    and $m_0 = 100$~GeV with (left) $\tanb = 10$ and
    (right) $\tan \beta = 55$.
    In the blue shaded regions the LSP is a sneutrino, and the other
    shadings and contours have the same meanings as in \reffig{m12m0}.
    }
  \label{fig:mumA300100planes}
\end{figure*}

As discussed earlier, in the NUHM2 we are free to adjust both soft Higgs
masses independently and as a consequence we can study models for which
both $\mu$ and $m_A$ are free parameters~\cite{Ajaib:2011ab}.
We first consider some $(\mu, m_A)$ planes for $A_0 = 0, m_{1/2} = 300$~GeV
and $m_0 = 100$~GeV as shown in \reffig{mumA300100planes}.
In each case, we see a WMAP-compatible rapid-annihilation funnel at
$m_A \sim 250$~GeV as the neutralino mass is roughly 0.43~$m_{1/2}$ and
therefore roughly half the pseudoscalar mass across the plane.
In addition there is an arc at relatively large $\mu$ and $m_A$ due to
the coannihilations between the neutralino and sneutrinos. The dark blue
shaded regions have a sneutrino LSP and are excluded~\cite{snu}.
In the left panel for $\tan \beta = 10$, we also see a near-vertical 
strip at low $\mu$ where the LSP has a substantial Higgsino component,
which extends downwards in a stau coannihilation strip almost parallel
to the charged dark LSP boundary. The $\mu > 0$ half of the plane is
consistent with the $g-2$ constraint. As we have seen before,
equilibrium is in this case well established along the vertical
transition strip, partially established along the horizontal funnel,
and not at all along the arcing coannihilation strip. While scattering
is heavily dominated by the spin-dependent cross-section along the
transition strip, spin-independent scattering is significant along the
other two strips.
In the right panel for $\tan \beta = 55$, the charged LSP boundary is
present at lower $\mu$ and more prevalent at large $m_A$. Both the stau
and sneutino LSP regions are again parallelled by WMAP-compatible
coannihilation strips. The supersymmetric contribution to $g-2$ in the
right panel exceeds the 2$\sigma$ bound.

In the left plane, the only region with an IceCube/DeepCore-friendly
neutrino flux is the vertical strip with a substantial Higgsino content,
analogous to those seen previously in our NUHM1 analysis. 
Here fluxes are in excess of 500~/km$^2$/yr.  On the other hand, in the
right panel where this strip has disappeared, the only
IceCube/DeepCore-friendly region is at small $\mu$ and $m_A$, below the
rapid-annihilation funnel. We further note that, in both planes, the
Higgs mass falls below 114~GeV.

\begin{figure*}
  \insertdoublefig{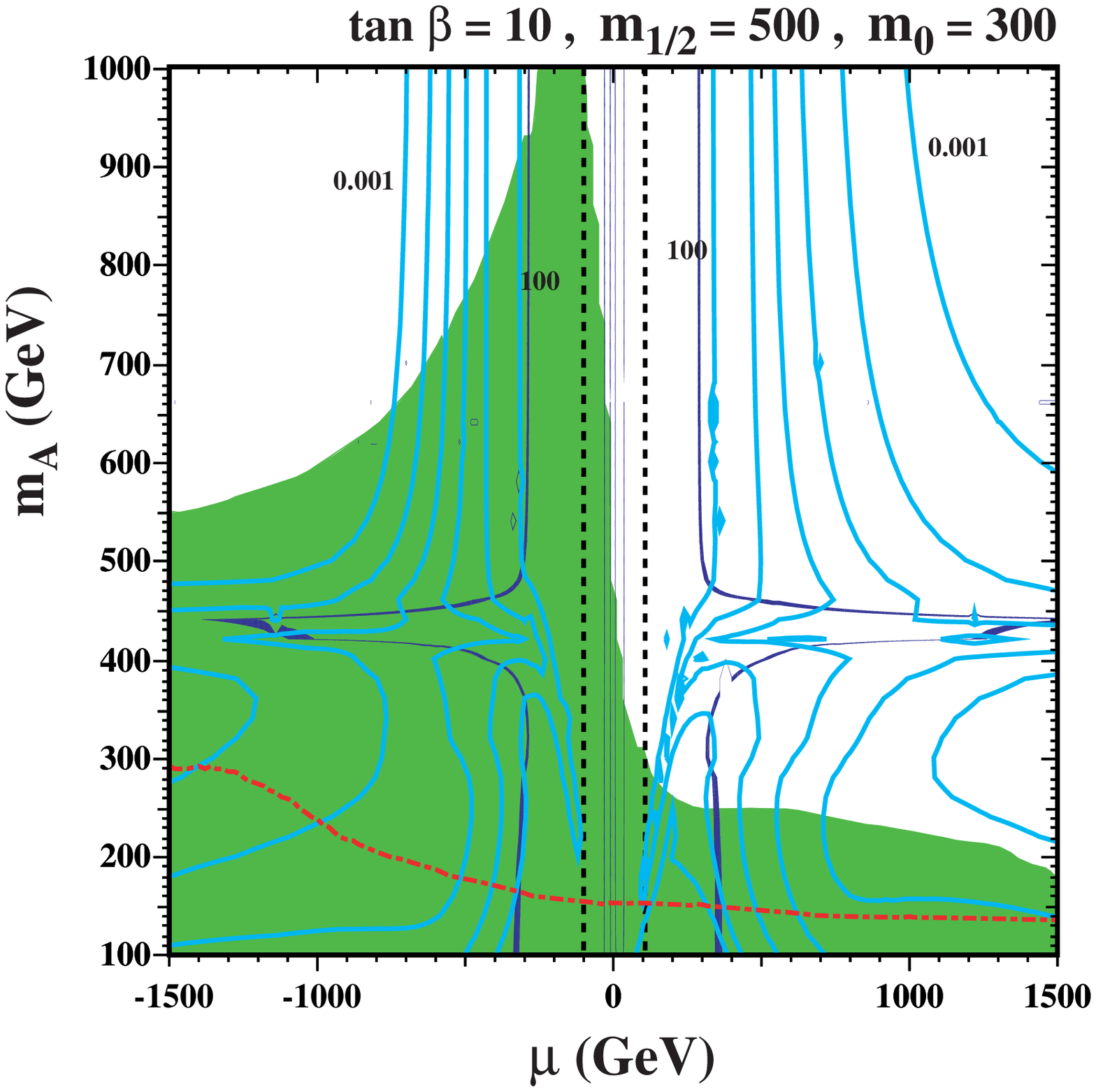}{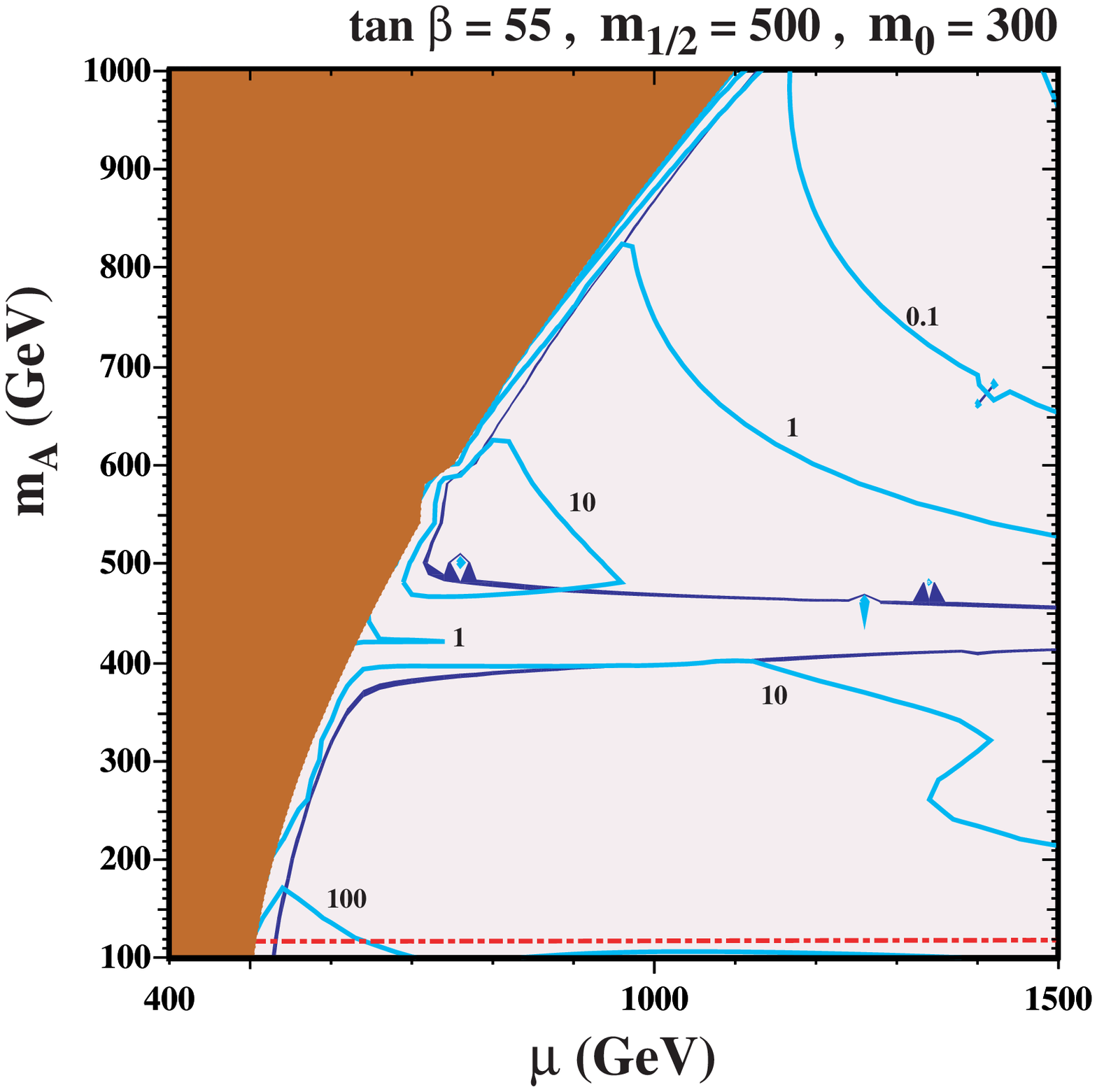}
  \caption{\it
    \coloronlinestatement
    The $(\mu, m_0)$ planes in the NUHM2 for $A_0 = 0, m_{1/2} = 500$~GeV
    and $m_0 = 300$~GeV with (left) $\tanb = 10$ and
    (right) $\tan \beta = 55$.
    The shadings and contours have the same meanings as in
    \reffig{m12m0}.
    }
  \label{fig:mumA500300planes}
\end{figure*}

Similar trends can be seen in \reffig{mumA500300planes}
for $A_0 = 0, m_{1/2} = 500$~GeV and $m_0 = 300$~GeV.
In the left panel for $\tan \beta = 10$, we see a near-vertical
strip where the LSP has an enhanced Higgsino component,
split in two by a near-horizontal rapid-annihilation funnel.
This funnel is also present in the right panel for $\tan \beta = 55$,
and splits a diagonal coannihilation strip close to the charged LSP
boundary. Again, as in the previous case for smaller $m_{1/2}$ and
$m_0$, the only IceCube/DeepCore-friendly regions are in the
Higgsino-like strip at small $\mu$ for $\tan \beta = 10$, and for very
small $\mu$ and $m_A$ for $\tan \beta = 55$. Spin-independent scattering
becomes dominant for $|\mu| \gae 500$~GeV when $\tan \beta = 10$ and is
dominant everywhere in the displayed plane for $\tan \beta = 55$.  In
this case, the Higgs mass is above 114~GeV everywhere above the red dot
dashed curved found at low $m_A$.  For $\tan \beta = 10$, the
supersymmetric contribution to $g-2$ is too small across the plane,
whereas for $\tan \beta = 55$ it lies within the experimental range.

\begin{figure*}
  \insertdoublefig{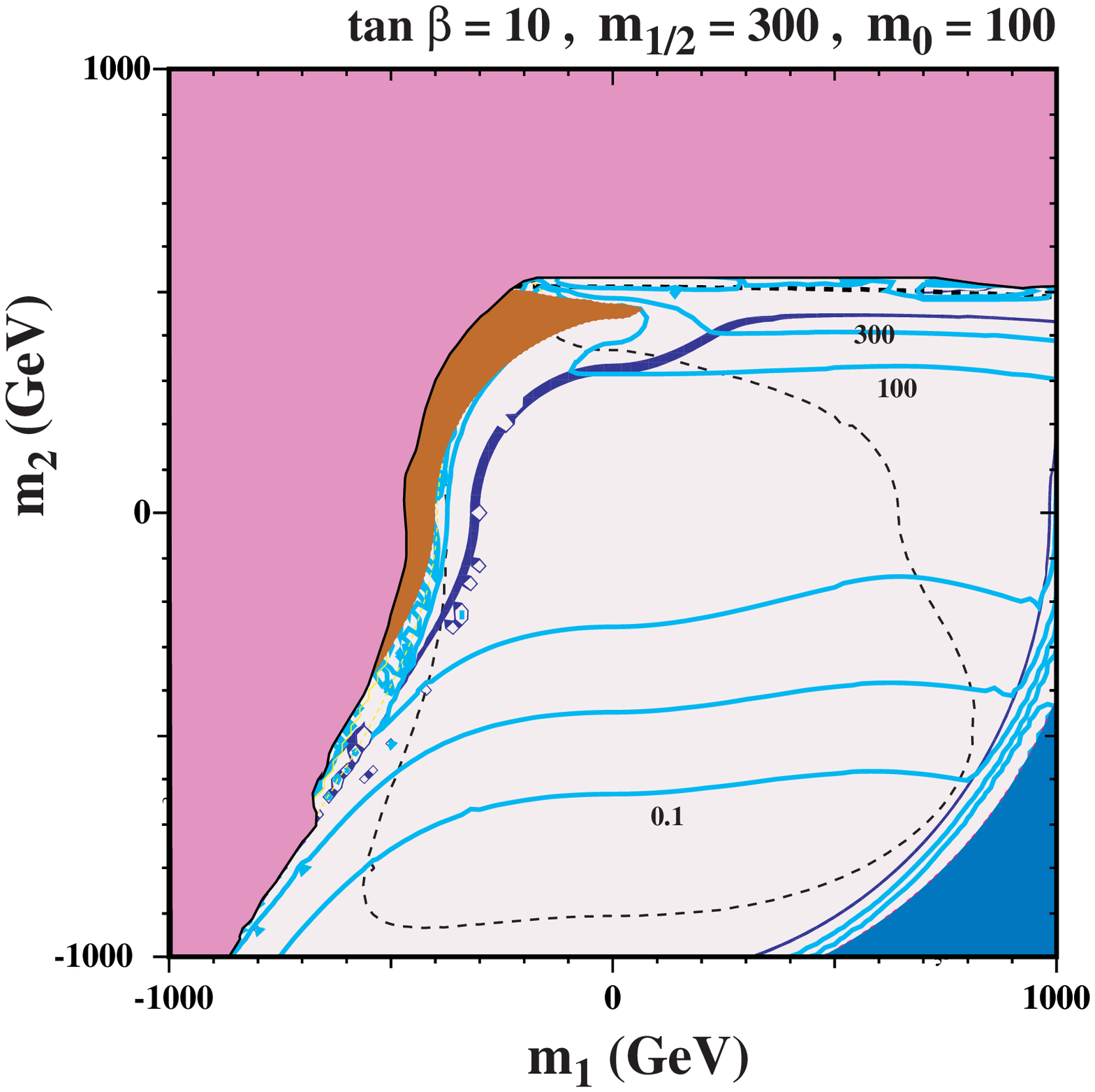}{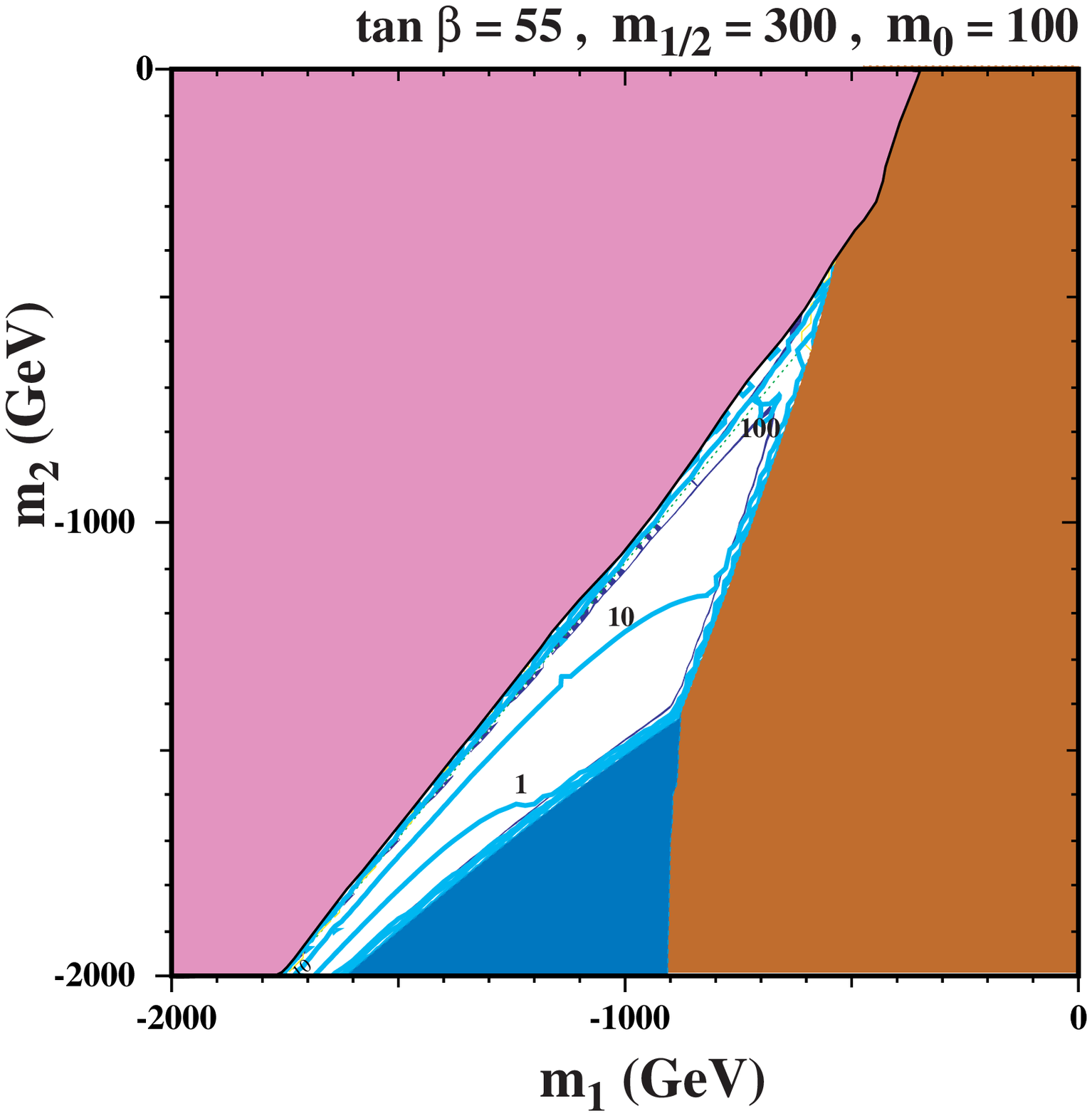}
  \caption{\it
    \coloronlinestatement
    The $(m_1, m_2)$ planes in the NUHM2 for $m_{1/2} = 300$~GeV
    and $m_0 = 100$~GeV with (left) $\tanb = 10$ and
    (right) $\tan \beta = 55$.
    The signs shown for $m_i$ actually refer to the sign of $m_i^2$
    as they are run in the RGEs.
    The shadings and contours have the same meanings as in
    \reffig{mumA300100planes}.
    }
  \label{fig:m1m2300100planes}
\end{figure*}

\begin{figure*}
  \insertdoublefig{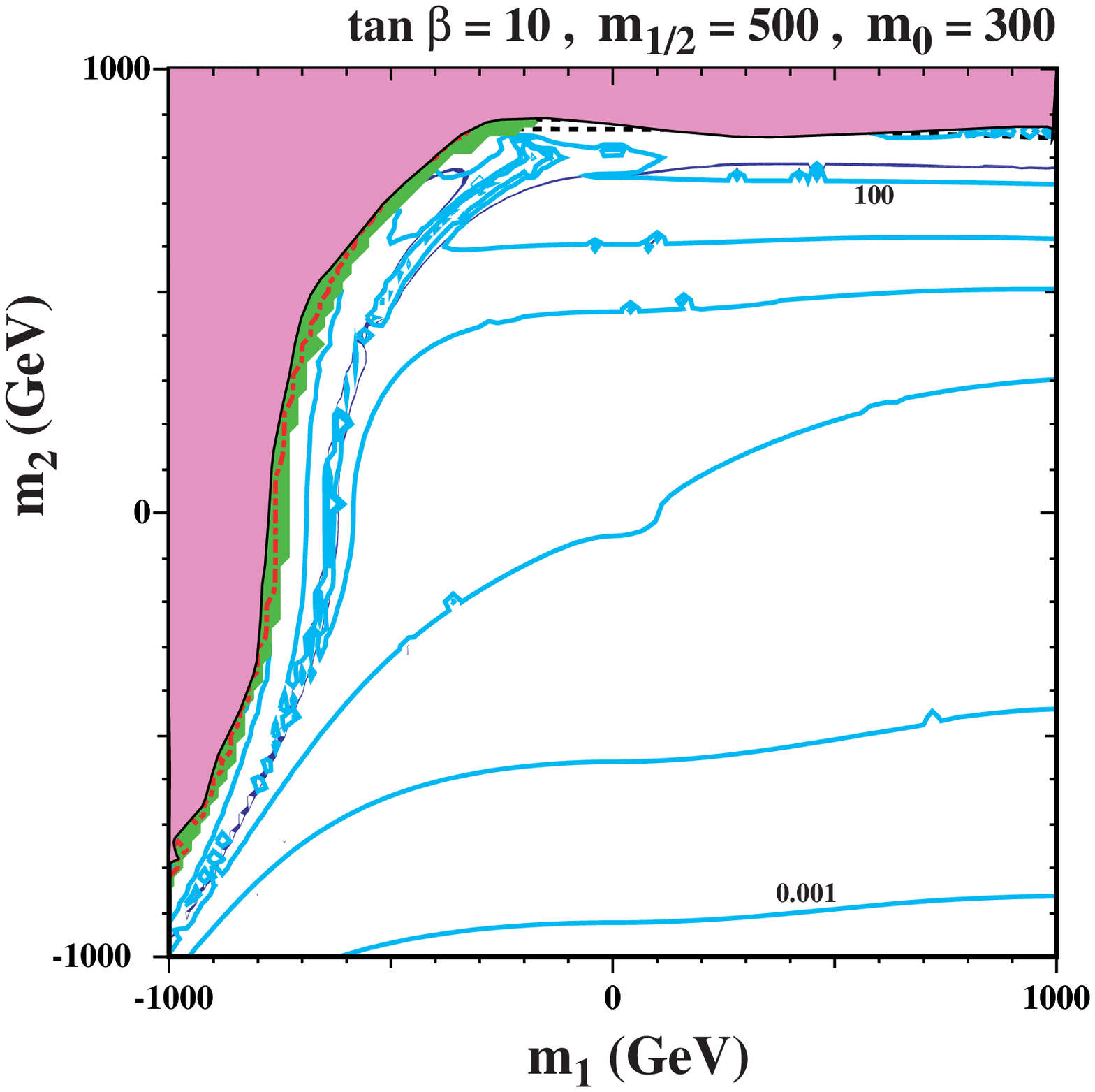}{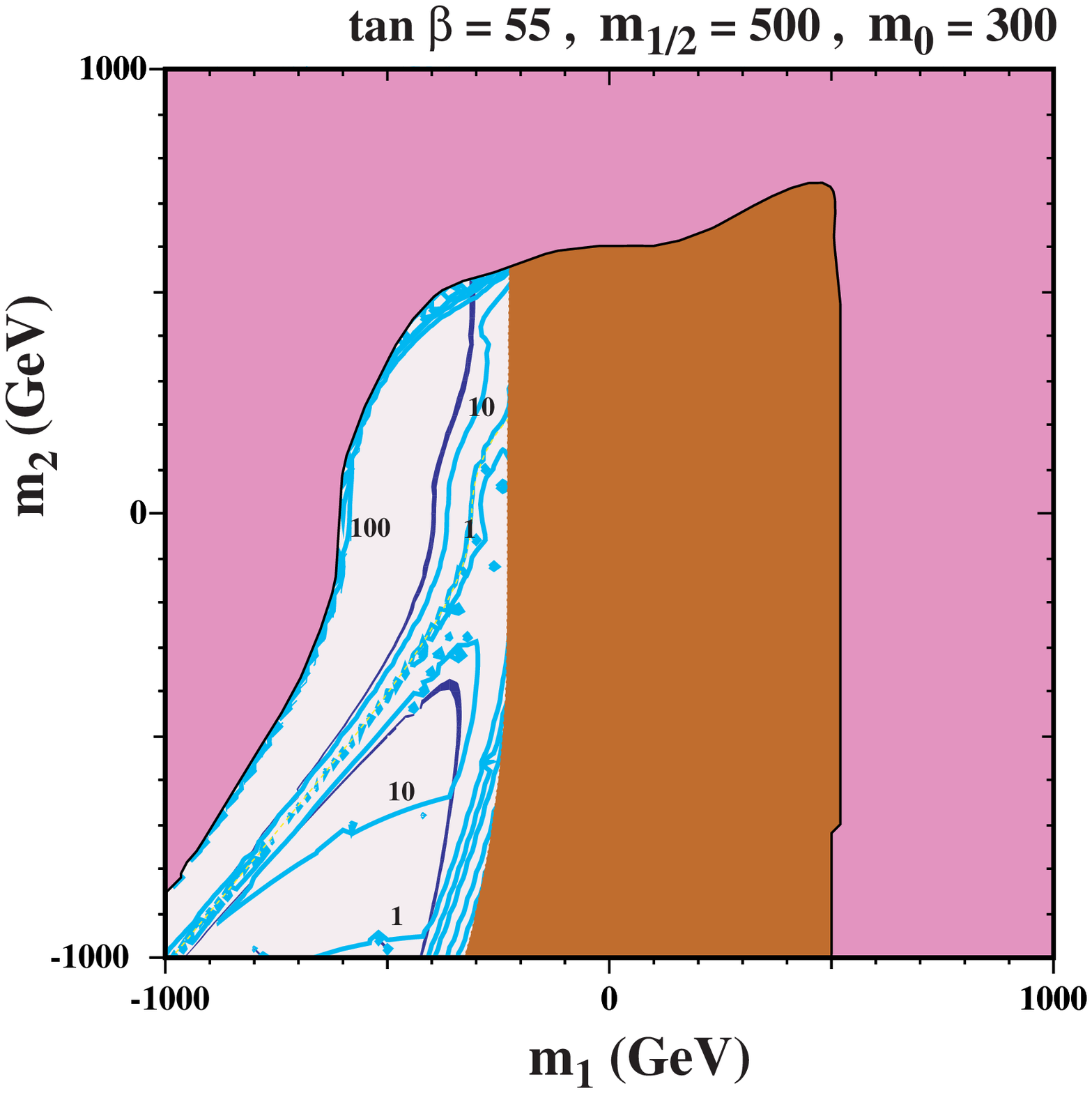}
  \caption{\it
    \coloronlinestatement
    The $(m_1, m_2)$ planes in the NUHM2 for $m_{1/2} = 500$~GeV
    and $m_0 = 300$~GeV with (left) $\tanb = 10$ and
    (right) $\tan \beta = 55$.
    The shadings and contours have the same meanings as in
    \reffig{m12m0}.
    }
  \label{fig:m1m2500300planes}
\end{figure*}

Finally, we present some $(m_1, m_2)$ planes in
Figs.~\ref{fig:m1m2300100planes} and~\ref{fig:m1m2500300planes}, for
$m_{1/2} = 300$~GeV, $m_0 = 100$~GeV and $m_{1/2} = 500$~GeV,
$m_0 = 300$~GeV, respectively. In each case, we assume $A_0 = 0$ and the
left panel is for $\tan \beta = 10$ and the right panel for
$\tan \beta = 55$.
The ranges of $m_1$ and $m_2$ are symmetric $\in (-1000, 1000)$~GeV,
except for the right panel of \reffig{m1m2300100planes}, where
asymmetric ranges $(m_1, m_2) \in (-2000, 0)$~GeV are chosen so as to
display better the regions not excluded by the EWSB constraint (pink)
and the sneutrino LSP regions (blue) or the charged-LSP constraint
(brown).
The signs shown for $m_i$ actually refer to the sign of $m_i^2$ as they
are run in the RGEs. As these are GUT scale parameters, negative values
may indicate a cosmological issue with broken symmetric vacua. For a
recent discussion of this issue in the NUHM, see \cite{Ellis:2008mc}.

In the left panel of \reffig{m1m2300100planes} for $\tan \beta = 10$,
we see two narrow WMAP strips following the top left and bottom right
boundaries of the allowed lozenge of the $(m_1, m_2)$ plane. Of these,
the former has an IceCube/DeepCore-friendly neutrino-induced muon flux
over most of its length, rising to $> 300$ events/km$^2$/yr for
$m_2 \sim 400$~GeV. Note that the nearly horizontal part of the strip
corresponds again to the transition strip where a there significant
Higgsino contribution to the neutralino composition.  The more vertical
part of the strip corresponds to the rapid annihilation funnel.  
On the other hand, the bottom right WMAP strip corresponding to
neutralino-sneutrino coannihilation generally has an unobservably small
neutrino flux, except for positive values of $m_2$, where the muon flux
may reach $30$ events/km$^2$/yr. As usual, the larger neutrino fluxes
are reached when the Higgsino component of the LSP is enhanced. In this
plane the $g-2$ constraint is satisfied, though the Higgs mass is low.

In the right panel of \reffig{m1m2300100planes} for $\tan \beta = 55$, 
the allowed region of the $(m_1, m_2)$ plane has receded to more
negative values of $m_1$ and $m_2$, and the two narrow WMAP strips are
squeezed closer together. Additionally, we note that the top left strip
has bifurcated along the two sides of a rapid-annihilation funnel,
whereas the lower right strip follows either the stau or sneutrino
coannihilation boundaries.
As in the left panel of \reffig{m1m2300100planes}, the neutrino fluxes
are generally more favourable along the top left strip, though they also
become more IceCube/DeepCore-friendly towards the upper end of the other
strip, near their junction. Neither $g-2$ or the Higgs mass constraints
are satisfied in this plane.

In the left panel of \reffig{m1m2500300planes} for $\tan \beta = 10$, we
see a single narrow WMAP strip following the top boundary of the allowed
region of the $(m_1, m_2)$ plane corresponding to the transition strip.
As in the corresponding panel of \reffig{m1m2300100planes}, the
neutrino-induced muon flux is largest, exceeding $100$ events/km$^2$/yr
and hence quite IceCube/DeepCore-friendly, in the top part of the strip
close to the EWSB boundary where $m_2 \sim 800$~GeV and the LSP has an
enhanced Higgsino component.
The left boundary corresponds to the the funnel region where there are
two strips, one of which continues on to the vertical transitions strip.
The upper part of the funnel still has observable fluxes, but these
quickly drop as $m_2$ is decreased. Here, and in the right panel as
well, equilibrium is established along all WMAP strips, however
spin-dependent scattering is sub-dominant almost everywhere in the plane.

In the right panel of \reffig{m1m2500300planes} for $\tan \beta = 55$, 
the previous single narrow WMAP strip has again bifurcated along the two 
sides of a rapid-annihilation funnel, and has moved away from the
boundaries of the allowed region of the $(m_1, m_2)$ plane. Once again,
the neutrino flux is largest for $m_2 > 0$, though generally lower than
in the left panel of \reffig{m1m2500300planes} (where $\tan \beta$ is
smaller) or in the right panel of \reffig{m1m2300100planes} (where
$m_{1/2}$ and $m_0$ are smaller), and only barely
IceCube/DeepCore-friendly. In this plane, both the $g-2$ and Higgs mass
constraints are satisfied.

\section{\label{sec:summary} Summary}

In our previous analysis of the CMSSM, we found that the flux of
high-energy neutrinos from LSP annihilations inside the Sun was likely
to be observable along the focus-point WMAP strip, where the Higgsino
component of the LSP is enhanced, and at the tip of the coannihilation
WMAP strip where the LSP is relatively light. On the other hand, there
were significant portions of the WMAP-compatible strips in parameter
space, particularly along the coannihilation strip and heavy-Higgs
rapid-annihilation funnels, where the the neutrino flux was not
IceCube/DeepCore-friendly.

We find some similar features in our analyses of the NUHM1 and NUHM2.
Specifically, there are significant portions of the WMAP strips where
the high-energy solar neutrino flux is unlikely to be observable with
IceCube/DeepCore. In these models, IceCube/DeepCore-friendly fluxes are
often found in regions where the LSP has an enhanced Higgsino component,
and this occurs under circumstances that cannot be realized in the
CMSSM. Specifically, it may occur for larger LSP masses than along the
focus-point strip of the CMSSM: see, for example, the right panel of
\reffig{m12m0}, the first three panels  of \reffig{mum12planes}, and the
first three panels of \reffig{mum0planes}.

We conclude, therefore, that IceCube/DeepCore has interesting prospects
for probing aspects of the NUHM1 and NUHM2 parameter spaces.
However, it seems clear that a more complete exploration of these
models, capable of measuring a high-energy neutrino flux when
the LSP is relatively heavy and/or does not have a large Higgsino
component, would require a subsequent generation of experiment.
On the other hand, one may hope that forthcoming LHC results and/or
direct searches for LSP scattering could provide more encouraging
indications on the prospects for searches for supersymmetric dark 
matter via annihilation into high-energy solar neutrinos.


\begin{acknowledgments}
  The work of KAO was supported in part by DOE Grant
  No.\ DE-FG02-94ER-40823. KAO also thanks SLAC 
  (supported by the DOE under contract number DE-AC02-76SF00515) and 
  the Stanford Institute for Theoretical Physics
  for their hospitality and support while this work was being finished.
  C.S.\ is grateful for financial support from the Swedish Research
  Council (VR) through the Oskar Klein Centre
  and thanks the William I.\ Fine Theoretical Physics Institute at the
  University of Minnesota, where part of this work was performed,
  for its hospitality.
  C.S.\ also thanks M.\ Danninger for useful discussions regarding
  IceCube/DeepCore.
  The work of V.C.S.\ was supported by Marie Curie International
  Reintegration grant SUSYDM-PHEN, MIRG-CT-2007-203189.
\end{acknowledgments}





\begin{thebibliography}{99}


\bibitem{indirectdet:solar}
  J.~Silk, K.~A.~Olive and M.~Srednicki,
  Phys.\ Rev.\ Lett.\  {\bf 55}, 257 (1985);
  M.~Srednicki, K.~A.~Olive and J.~Silk,
  Nucl.\ Phys.\  B {\bf 279}, 804 (1987);
  J.~S.~Hagelin, K.~W.~Ng and K.~A.~Olive,
  Phys.\ Lett.\  B {\bf 180}, 375 (1986);
  K.~W.~Ng, K.~A.~Olive and M.~Srednicki,
  Phys.\ Lett.\  B {\bf 188}, 138 (1987);
  T.~K.~Gaisser, G.~Steigman and S.~Tilav,
  Phys.\ Rev.\  D {\bf 34}, 2206 (1986);
  F.~Halzen, T.~Stelzer and M.~Kamionkowski,
  Phys.\ Rev.\  D {\bf 45}, 4439 (1992);
  L.~Bergstrom, J.~Edsjo and P.~Gondolo,
  Phys.\ Rev.\  D {\bf 55}, 1765 (1997)
  [arXiv:hep-ph/9607237];
  K.~Freese and M.~Kamionkowski,
  Phys.\ Rev.\ D {\bf 55}, 1771 (1997)
  [arXiv:hep-ph/9609370].

\bibitem{indirectdet:earth}
  K.~Freese,
  Phys.\ Lett.\ B {\bf 167}, 295 (1986);
  L.~M.~Krauss, M.~Srednicki and F.~Wilczek,
  Phys.\ Rev.\ D {\bf 33}, 2079 (1986).

\bibitem{Ahrens:2003ix}
  J.~Ahrens {\it et al.}  [IceCube Collaboration],
  Astropart.\ Phys.\  {\bf 20}, 507 (2004)
  [arXiv:astro-ph/0305196].

\bibitem{Achterberg:2006md}
  A.~Achterberg {\it et al.}  [IceCube Collaboration],
  Astropart.\ Phys.\  {\bf 26}, 155 (2006)
  [arXiv:astro-ph/0604450].

\bibitem{Resconi:2008fe}
  E.~Resconi, for the IceCube Collaboration,
  Nucl.\ Instrum.\ Meth.\  A {\bf 602}, 7 (2009)
  [arXiv:0807.3891 [astro-ph]].

\bibitem{Abbasi:2009uz}
  R.~Abbasi {\it et al.}  [IceCube Collaboration],
  Phys.\ Rev.\ Lett.\  {\bf 102}, 201302 (2009)
  [arXiv:0902.2460v2 [astro-ph.CO]];
  the combined IceCube/DeepCore flux sensitivities are found in
  Figure~3 of the arXiv version~2.

\bibitem{EOSSnu}
  J.~Ellis, K.~A.~Olive, C.~Savage and V.~C.~Spanos,
  Phys.\ Rev.\  D {\bf 81}, 085004 (2010)
  [arXiv:0912.3137 [hep-ph]].

\bibitem{cmssm}
  M.~Drees and M.~M.~Nojiri,
  Phys.\ Rev.\  D {\bf 47}, 376 (1993)
  [arXiv:hep-ph/9207234];
  H.~Baer and M.~Brhlik,
  Phys.\ Rev.\  D {\bf 53}, 597 (1996)
  [arXiv:hep-ph/9508321];
  H.~Baer and M.~Brhlik,
  Phys.\ Rev.\  D {\bf 57}, 567 (1998)
  [arXiv:hep-ph/9706509];
  J.~R.~Ellis, T.~Falk, K.~A.~Olive and M.~Schmitt,
  Phys.\ Lett.\ B {\bf 388}, 97 (1996)
  [arXiv:hep-ph/9607292];
  J.~R.~Ellis, T.~Falk, K.~A.~Olive and M.~Schmitt,
  Phys.\ Lett.\  B {\bf 413}, 355 (1997)
  [arXiv:hep-ph/9705444];
  J.~R.~Ellis, T.~Falk, G.~Ganis, K.~A.~Olive and M.~Schmitt,
  Phys.\ Rev.\  D {\bf 58}, 095002 (1998)
  [arXiv:hep-ph/9801445];
  V.~D.~Barger and C.~Kao,
  Phys.\ Rev.\  D {\bf 57}, 3131 (1998)
  [arXiv:hep-ph/9704403];
  J.~R.~Ellis, T.~Falk, G.~Ganis and K.~A.~Olive,
  Phys.\ Rev.\  D {\bf 62}, 075010 (2000)
  [arXiv:hep-ph/0004169];
  J.~R.~Ellis, T.~Falk, G.~Ganis, K.~A.~Olive and M.~Srednicki,
  Phys.\ Lett.\  B {\bf 510}, 236 (2001)
  [arXiv:hep-ph/0102098];
  V.~D.~Barger and C.~Kao,
  Phys.\ Lett.\  B {\bf 518}, 117 (2001)
  [arXiv:hep-ph/0106189];
  L.~Roszkowski, R.~Ruiz de Austri and T.~Nihei,
  JHEP {\bf 0108}, 024 (2001)
  [arXiv:hep-ph/0106334];
  A.~B.~Lahanas and V.~C.~Spanos,
  Eur.\ Phys.\ J.\  C {\bf 23}, 185 (2002)
  [arXiv:hep-ph/0106345];
  A.~Djouadi, M.~Drees and J.~L.~Kneur,
  JHEP {\bf 0108}, 055 (2001)
  [arXiv:hep-ph/0107316];
  U.~Chattopadhyay, A.~Corsetti and P.~Nath,
  Phys.\ Rev.\  D {\bf 66}, 035003 (2002)
  [arXiv:hep-ph/0201001];
  J.~R.~Ellis, K.~A.~Olive and Y.~Santoso,
  New J.\ Phys.\  {\bf 4}, 32 (2002)
  [arXiv:hep-ph/0202110];
  H.~Baer, C.~Balazs, A.~Belyaev, J.~K.~Mizukoshi, X.~Tata and Y.~Wang,
  JHEP {\bf 0207}, 050 (2002)
  [arXiv:hep-ph/0205325];
  R.~Arnowitt and B.~Dutta,
  arXiv:hep-ph/0211417.

\bibitem{Komatsu:2010fb}
  E.~Komatsu {\it et al.}  [WMAP Collaboration],
  Astrophys.\ J.\ Suppl.\  {\bf 192} (2011) 18
  [arXiv:1001.4538 [astro-ph.CO]].

\bibitem{eoss}
  J.~R.~Ellis, K.~A.~Olive, Y.~Santoso and V.~C.~Spanos,
  Phys.\ Lett.\  B {\bf 565}, 176 (2003)
  [arXiv:hep-ph/0303043];

\bibitem{cmssmwmap}
  H.~Baer and C.~Balazs,
  JCAP {\bf 0305}, 006 (2003)
  [arXiv:hep-ph/0303114];
  A.~B.~Lahanas and D.~V.~Nanopoulos,
  Phys.\ Lett.\  B {\bf 568}, 55 (2003)
  [arXiv:hep-ph/0303130];
  U.~Chattopadhyay, A.~Corsetti and P.~Nath,
  Phys.\ Rev.\  D {\bf 68}, 035005 (2003)
  [arXiv:hep-ph/0303201];
  C.~Munoz,
  Int.\ J.\ Mod.\ Phys.\  A {\bf 19}, 3093 (2004)
  [arXiv:hep-ph/0309346];
  R.~L.~Arnowitt, B.~Dutta and B.~Hu,
  arXiv:hep-ph/0310103.

\bibitem{cmssmnu}
  A.~Corsetti and P.~Nath,
  Int.\ J.\ Mod.\ Phys.\  A {\bf 15}, 905 (2000)
  [arXiv:hep-ph/9904497];
  J.~L.~Feng, K.~T.~Matchev and F.~Wilczek,
  Phys.\ Rev.\  D {\bf 63}, 045024 (2001)
  [arXiv:astro-ph/0008115];
  V.~D.~Barger, F.~Halzen, D.~Hooper and C.~Kao,
  Phys.\ Rev.\  D {\bf 65}, 075022 (2002)
  [arXiv:hep-ph/0105182];
  H.~Baer, A.~Belyaev, T.~Krupovnickas and J.~O'Farrill,
  JCAP {\bf 0408}, 005 (2004)
  [arXiv:hep-ph/0405210];
  R.~Trotta, R.~R.~de Austri and C.~P.~d.~Heros,
  JCAP {\bf 0908}, 034 (2009)
  [arXiv:0906.0366 [astro-ph.HE]].

\bibitem{nuhm1}
  H.~Baer, A.~Mustafayev, S.~Profumo, A.~Belyaev and X.~Tata,
  Phys.\ Rev.\  D {\bf 71}, 095008 (2005)
  [arXiv:hep-ph/0412059];
  H.~Baer, A.~Mustafayev, S.~Profumo, A.~Belyaev and X.~Tata,
  JHEP {\bf 0507}, 065 (2005)
  [arXiv:hep-ph/0504001].

\bibitem{nuhm12}
  J.~R.~Ellis, K.~A.~Olive and P.~Sandick,
  Phys.\ Rev.\  D {\bf 78}, 075012 (2008)
  [arXiv:0805.2343 [hep-ph]].

\bibitem{nonu}
  D.~Matalliotakis and H.~P.~Nilles,
  Nucl.\ Phys.\  B {\bf 435}, 115 (1995)
  [arXiv:hep-ph/9407251];
  M.~Olechowski and S.~Pokorski,
  Phys.\ Lett.\  B {\bf 344}, 201 (1995)
  [arXiv:hep-ph/9407404];
  V.~Berezinsky, A.~Bottino, J.~R.~Ellis, N.~Fornengo, G.~Mignola and
  S.~Scopel,
  Astropart.\ Phys.\  {\bf 5}, 1 (1996)
  [arXiv:hep-ph/9508249];
  M.~Drees, M.~M.~Nojiri, D.~P.~Roy and Y.~Yamada,
  Phys.\ Rev.\  D {\bf 56}, 276 (1997)
  [Erratum-ibid.\  D {\bf 64}, 039901 (2001)]
  [arXiv:hep-ph/9701219];
  M.~Drees, Y.~G.~Kim, M.~M.~Nojiri, D.~Toya, K.~Hasuko and T.~Kobayashi,
  Phys.\ Rev.\  D {\bf 63}, 035008 (2001)
  [arXiv:hep-ph/0007202];
  P.~Nath and R.~L.~Arnowitt,
  Phys.\ Rev.\  D {\bf 56}, 2820 (1997)
  [arXiv:hep-ph/9701301];
  J.~R.~Ellis, T.~Falk, G.~Ganis, K.~A.~Olive and M.~Schmitt,
  Phys.\ Rev.\  D {\bf 58}, 095002 (1998)
  [arXiv:hep-ph/9801445];
  J.~R.~Ellis, T.~Falk, G.~Ganis and K.~A.~Olive,
  Phys.\ Rev.\  D {\bf 62}, 075010 (2000)
  [arXiv:hep-ph/0004169];
  A.~Bottino, F.~Donato, N.~Fornengo and S.~Scopel,
  Phys.\ Rev.\  D {\bf 63}, 125003 (2001)
  [arXiv:hep-ph/0010203];
  S.~Profumo,
  Phys.\ Rev.\  D {\bf 68}, 015006 (2003)
  [arXiv:hep-ph/0304071];
  D.~G.~Cerdeno and C.~Munoz,
  JHEP {\bf 0410}, 015 (2004)
  [arXiv:hep-ph/0405057].

\bibitem{nuhm2}
  J.~R.~Ellis, K.~A.~Olive and Y.~Santoso,
  Phys.\ Lett.\  B {\bf 539}, 107 (2002)
  [arXiv:hep-ph/0204192];
  J.~R.~Ellis, T.~Falk, K.~A.~Olive and Y.~Santoso,
  Nucl.\ Phys.\  B {\bf 652}, 259 (2003)
  [arXiv:hep-ph/0210205].

\bibitem{Ahmed:2009zw}
  Z.~Ahmed {\it et al.}  [The CDMS-II Collaboration],
  Science {\bf 327}, 1619 (2010)
  [arXiv:0912.3592 [astro-ph.CO]].

\bibitem{Aprile:2010um}
  E.~Aprile {\it et al.}  [XENON100 Collaboration],
  Phys.\ Rev.\ Lett.\  {\bf 105}, 131302 (2010)
  [arXiv:1005.0380 [astro-ph.CO]].

\bibitem{CMSsusy}
  V.~Khachatryan {\it et al.}  [CMS Collaboration],
  Phys.\ Lett.\  B {\bf 698}, 196 (2011)
  [arXiv:1101.1628 [hep-ex]].

\bibitem{ATLASsusy}
  G.~Aad {\it et al.}  [ATLAS Collaboration],
  in preparation;
  see \\
  \url{https://twiki.cern.ch/twiki/bin/view/AtlasPublic/SusyPublicResults}.

\bibitem{mc3}
  O.~Buchmueller {\it et al.},
  JHEP {\bf 0809}, 117 (2008)
  [arXiv:0808.4128 [hep-ph]];
  O.~Buchmueller {\it et al.},
  Eur.\ Phys.\ J.\  C {\bf 64}, 391 (2009)
  [arXiv:0907.5568 [hep-ph]].

\bibitem{mcmSUGRA}
  O.~Buchmueller {\it et al.},
  Eur.\ Phys.\ J.\  C {\bf 71}, 1583 (2011)
  [arXiv:1011.6118 [hep-ph]].

\bibitem{bsgex}
  S.~Chen {\it et al.}  [CLEO Collaboration],
  Phys.\ Rev.\ Lett.\  {\bf 87}, 251807 (2001)
  [arXiv:hep-ex/0108032];
  P.~Koppenburg {\it et al.}  [Belle Collaboration],
  Phys.\ Rev.\ Lett.\  {\bf 93}, 061803 (2004)
  [arXiv:hep-ex/0403004];
  B.~Aubert {\it et al.}  [BaBar Collaboration],
    arXiv:hep-ex/0207076;
  E.~Barberio {\it et al.}  [Heavy Flavor Averaging Group (HFAG)],
  arXiv:hep-ex/0603003.

\bibitem{LEPsusy}
  Joint LEP~2 Supersymmetry Working Group,
  {\it Combined LEP Chargino Results up to 208 GeV},
  \url{http://lepsusy.web.cern.ch/lepsusy/www/inos_moriond01/charginos_pub.html}.

\bibitem{LEPHiggs}
  R.~Barate {\it et al.}  [ALEPH, DELPHI, L3, OPAL Collaborations:
  the LEP Working Group for Higgs boson searches],
  Phys.\ Lett.\  B {\bf 565}, 61 (2003)
  [arXiv:hep-ex/0306033];
  D.~Zer-Zion,
  {\it Prepared for 32nd International Conference on High-Energy
       Physics (ICHEP 04),
       Beijing, China, 16-22 Aug 2004},
  World Scientific, Hackensack (2005);
  ALEPH, DELPHI, L3, OPAL Collaborations:
  the LEP Working Group for Higgs boson searches,
  LHWG-NOTE-2004-01, ALEPH-2004-008, DELPHI-2004-042, L3-NOTE-2820,
  OPAL-TN-744,
  \url{http://lephiggs.web.cern.ch/LEPHIGGS/papers/August2004_MSSM/index.html}.

\bibitem{FeynHiggs}
  S.~Heinemeyer, W.~Hollik and G.~Weiglein,
  Comput.\ Phys.\ Commun.\  {\bf 124}, 76 (2000)
  [arXiv:hep-ph/9812320];
  S.~Heinemeyer, W.~Hollik and G.~Weiglein,
  Eur.\ Phys.\ J.\  C {\bf 9}, 343 (1999)
  [arXiv:hep-ph/9812472];
  G.~Degrassi, S.~Heinemeyer, W.~Hollik, P.~Slavich and G.~Weiglein,
  Eur.\ Phys.\ J.\  C {\bf 28}, 133 (2003)
  [arXiv:hep-ph/0212020];
  M.~Frank, T.~Hahn, S.~Heinemeyer, W.~Hollik, H.~Rzehak
  and G.~Weiglein,
  JHEP {\bf 0702}, 047 (2007)
  [arXiv:hep-ph/0611326];
  \url{http://www.feynhiggs.de/}

\bibitem{Bennett:2006fi}
  G.~W.~Bennett {\it et al.}  [Muon G-2 Collaboration],
  Phys.\ Rev.\  D {\bf 73}, 072003 (2006)
  [arXiv:hep-ex/0602035].

\bibitem{g-2babar}
  B.~Aubert {\it et al.}  [BABAR Collaboration],
  Phys.\ Rev.\ Lett.\  {\bf 103}, 231801 (2009)
  [arXiv:0908.3589 [hep-ex]].

\bibitem{newg-2} 
  M.~Davier, A.~Hoecker, B.~Malaescu and Z.~Zhang,
  Eur.\ Phys.\ J.\  C {\bf 71}, 1515 (2011)
  [arXiv:1010.4180 [hep-ph]].

\bibitem{Bottino:1999ei}
  A.~Bottino, F.~Donato, N.~Fornengo and S.~Scopel,
  Astropart.\ Phys.\  {\bf 13}, 215 (2000)
  [arXiv:hep-ph/9909228].

\bibitem{Accomando:1999eg}
  E.~Accomando, R.~L.~Arnowitt, B.~Dutta and Y.~Santoso,
  Nucl.\ Phys.\  B {\bf 585}, 124 (2000)
  [arXiv:hep-ph/0001019].

\bibitem{Ellis:2005mb}
  J.~R.~Ellis, K.~A.~Olive, Y.~Santoso and V.~C.~Spanos,
  Phys.\ Rev.\  D {\bf 71}, 095007 (2005)
  [arXiv:hep-ph/0502001].

\bibitem{Ellis:2008hf}
  J.~R.~Ellis, K.~A.~Olive and C.~Savage,
  Phys.\ Rev.\  D {\bf 77}, 065026 (2008)
  [arXiv:0801.3656 [hep-ph]].

\bibitem{Niro:2009mw}
  V.~Niro, A.~Bottino, N.~Fornengo and S.~Scopel,
  Phys.\ Rev.\  D {\bf 80}, 095019 (2009)
  [arXiv:0909.2348 [hep-ph]].

\bibitem{Alekseev:2007vi}
  M.~Alekseev {\it et al.}  [COMPASS Collaboration],
  Phys.\ Lett.\  B {\bf 660}, 458 (2008)
  [arXiv:0707.4077 [hep-ex]].

\bibitem{Serenelli:2009yc}
  A.~Serenelli, S.~Basu, J.~W.~Ferguson and M.~Asplund,
  Astrophys.\ J.\  {\bf 705}, L123 (2009)
  [arXiv:0909.2668 [astro-ph.SR]].

\bibitem{Asplund:2009fu}
  M.~Asplund, N.~Grevesse, A.~J.~Sauval and P.~Scott,
  Ann.\ Rev.\ Astron.\ Astrophys.\  {\bf 47}, 481 (2009)
  [arXiv:0909.0948 [astro-ph.SR]].

\bibitem{Gould:1992xx}
  A.~Gould,
  Astrophys.\ J.\  {\bf 388}, 338 (1992).

\bibitem{Barger:2011em}
  V.~Barger, Y.~Gao and D.~Marfatia,
  Phys.\ Rev.\  D {\bf 83}, 055012 (2011)
  [arXiv:1101.4410 [hep-ph]].

\bibitem{Danninger:2011pc}
  M.~Danninger,
  private correspondence.

\bibitem{wimpsim}
  M.~Blennow, J.~Edsjo and T.~Ohlsson,
  JCAP {\bf 0801}, 021 (2008)
  [arXiv:0709.3898 [hep-ph]];
  J.~Edsjo,
  WimpSim Neutrino Monte Carlo,
  \url{http://www.physto.se/~edsjo/wimpsim/}

\bibitem{darksusy}
  P.~Gondolo, J.~Edsjo, P.~Ullio, L.~Bergstrom, M.~Schelke and
  E.~A.~Baltz,
  JCAP {\bf 0407}, 008 (2004)
  [arXiv:astro-ph/0406204];
  P.~Gondolo, J.~Edsjo, P.~Ullio, L.~Bergstrom, M.~Schelke,
  E.~A.~Baltz, T.~Bringmann and G.~Duda,
  \url{http://www.physto.se/~edsjo/darksusy/}

\bibitem{Ajaib:2011ab}
  For a recent study within the NUHM2, see:
  M.~A.~Ajaib, I.~Gogoladze and Q.~Shafi,
  arXiv:1101.0835 [hep-ph].

\bibitem{snu}
  T.~Falk, K.~A.~Olive and M.~Srednicki,
  Phys.\ Lett.\  B {\bf 339}, 248 (1994)
  [arXiv:hep-ph/9409270];
  C.~Arina and N.~Fornengo,
  JHEP {\bf 0711}, 029 (2007)
  [arXiv:0709.4477 [hep-ph]].

\bibitem{Ellis:2008mc}
  J.~R.~Ellis, J.~Giedt, O.~Lebedev, K.~Olive and M.~Srednicki,
  Phys.\ Rev.\  D {\bf 78}, 075006 (2008)
  [arXiv:0806.3648 [hep-ph]].



\end{thebibliography}
\end{document}